\documentstyle[psfig]{mn}

\newcommand{\eq}{\begin{equation}}
\newcommand{\en}{\end{equation}}
\newcommand{\lab}{\label}

\begin{document}

%\onecolumn

\title[Pulsar emission]{An Empirical Model for the Radio Emission from Pulsars}

\author[Geoff Wright]{G.A.E.Wright\\
 Astronomy Centre, University of Sussex, Falmer, BN1 9QJ, UK}

\date{Accepted.........Received.....; in original form 2002...}

\maketitle

\begin{abstract}

A pulsar model is proposed which involves the entire magnetosphere in
the production of the observed coherent radio emission. The
observationally-inferred regularity of peaks in the pulsar profiles of
`slow' pulsars (Rankin 1990,1993a), is shown to suggest that inner and
outer cones of emission near the polar cap interact with and `mirror'
two rings in the outer magnetosphere: one where the null line
intersects the light-cylinder, and another where it intersects the
boundary of the corotating dead zone.  The observed dependency of
conal type on period is shown to follow naturally from the assumption
that cones only form when the mirror intersection points lie between
two fixed heights from the surface, suggesting that a feedback system
exists between the surface and the mirror points, accomplished by a
flow of charges of opposite sign in either direction. In their flow to
and from the mirror points, the particles execute an azimuthal drift
around the magnetic pole, thereby creating a ring of discrete
`emission nodes' close to the surface. Motion of the nodes is observed
as subpulse `drift', which is interpreted here as a small residual
component of the real particle drift. The nodes can move in either
direction or even remain stationary, and can differ in the inner and
outer cones. A precise fit is found for the drifting subpulses of
PSR0943+10. Azimuthal interactions between different regions of the
magnetosphere depend on the angle between the magnetic and rotation
axes and influence the conal type, as observed. The model sees `slow'
pulsars as being at the end of an evolutionary development where the
outer gap region no longer produces pair cascades, but is still the
intermittent source of low-energy pairs in a magnetosphere-wide
feedback system.

\end{abstract}

\section{INTRODUCTION}

In 1975 Ruderman $\&$ Sutherland (henceforth RS) proposed a model for
pulsar emission which postulated a gap region located immediately
above the neutron star surface and, as a result of the inability of
the electric field to remove ions from the surface, subject to an
intense potential difference of around $10^{14}eV$. In this gap
pair-production in the intense magnetic field generated electrons
which heated the surface and created localised discharging
regions. These `sparks' drifted around the magnetic pole and ejected
energetic particles into the magnetosphere where a bunching mechanism,
first proposed by Sturrock (1971), created coherent radiation
following a two-stream instability. The model has been highly
influential, since it provides a quantitative framework within which
theorists and observers can work, and one of its underlying tenets,
that the sources of pulsar radio emission are plasma columns
circulating just above the polar cap, is now widely accepted.

Nevertheless, over the years the polar gap model has been questioned
on theoretical grounds, largely through difficulties with the neutron
star surface binding energy (e.g. Jones 1985, 1986, Abraham $\&$
Shapiro, 1991, Neuhauser et al, 1987) and with the high plasma
densities required in the emisson regions (Lesch et al, 1998).  On
observational grounds too it is not easy to reconcile the model with
the complexity of subpulse behaviour, such as stationary or
counter-drifting subpulses (e.g. Biggs et al. 1985, Nowakowski 1991),
mode changing and nulling. Evidence that the emission profiles
generally take the form of nested cones (Rankin 1983, Lyne $\&$
Manchester 1988, Rankin 1990(RI), 1993a(RII), 1993b) has presented a
further challenge to the model, and only by appealing to surface
multipoles (e.g. Gil et al, 2002a,b) does it seem possible to confine
the `sparks' either radially or azimuthally. More recently, the
precisely drifting emission columns of PSR0943+10, analysed in great
detail by Deshpande $\&$ Rankin (1999, 2001 (henceforth DR)),
convincingly confirm the picture of circulating plasma
columns. However the observed circulation rate is only reconcilable
with the RS model by adopting a potential difference across the cap
which is significantly lower than that predicted by the RS model, and
- if generated by surface magnetic features - the columns would seem
to require multipoles with a high degree of regularity (Asseo $\&$
Khechinasvili 2001, Gil et al 2002a,b).

In response to these difficulties, some authors have analysed
magnetospheres based on free electron flows from the polar cap and
gradual acceleration into the upper magnetosphere (`inner accelerator'
models) (e.g. Arons $\&$ Scharlemann 1979, Mestel $\&$ Shibata 1994,
Hibschmann $\&$ Arons 2001a,b (henceforth HAa,b), Harding $\&$
Muslimov 2002 (HM), Harding et al 2002 (HMZ), Harding $\&$ Muslimov
2003).  Such models are often part of wider attempts at creating a
self-consistent global magnetosphere (Mestel et al 1985, Shibata 1995,
Mestel 1999 and references therein). Although arguably more consistent
with the physics of the neutron star surface - and with the overall
current balance and torque transfer requirements - these models lack
the predictive power of the RS model when faced with detailed radio
observations. Possibly the belief prevails that radio emission,
undoubtedly originating just a few tens of stellar radio above the
surface and energetically weak compared to the spin-down energy loss,
has little to say about global conditions. Furthermore the sheer
complexity of highly time-dependent subpulse phenomena acts as a great
deterrent to theoreticians seeking to define a steady state condition.

In trying to bring the large-scale magnetosphere models closer to
observational testing, and in a deliberate attempt to explore the link
between the the observed polar cap activity and the outer
magnetosphere, we have taken a fresh look at the radio profiles
analysed in RI and RII. Pulsar integrated profiles are suitable to
this purpose, since they are one feature of pulsar observations which
displays great stability: a pulsar's profile is its invariant
signature. Rankin's conal classification of the profile forms is here
interpreted essentially from a geometric standpoint without prejudice
for or against any particular formation process. In essence, we seek
to use the profiles as diagnostics of the magnetosphere's structure.

This paper argues that if regular conal structures exist in all or
most pulsar profiles - as claimed in RI and RII - then explanations in
terms of complex polar magnetic field topology are difficult to
support, since they would require similar magnetic topologies from
pulsar to pulsar.  However it is suggested here that conal emission
may be the natural geometric result of interactions between the polar
cap and outer regions of the outer magnetosphere in the purely dipolar
environment of `slow` older pulsars, provided that the once prolific
outergap production of pairs of a pulsar in its early life is now weak
and intermittent. This proposed evolutionary link between the younger
faster pulsars (such as the Crab pulsar), whose principle emission is
in the gamma-ray band, and the older weaker radio pulsars is seen as a
major and novel feature of the model here.

First, in section (2), it is argued that the fixed ratio of the cone
radii is consistent with the assumption that emission occurs
preferentially on two critical sets of field-lines, one bounding the
corotating dead zone (defined by the last field-line to close within
the light-cylinder), the other passing through the intersection of the
null line (defined as the boundary between regions of net charge
density of opposing sign) and the light cylinder. In section (3) it is
shown that the observed period dependence of conal types can be simply
explained if emission is possible only if a `mirror' point (i.e. an
intersection of the null line with the critical field-lines) lies
between two fixed altitudes. In section (4) it is shown how the
inevitable drift of the inflowing and outflowing particles about the
magnetic axis causes `emission nodes' to form above the surface, and,
in section (5), that this leads to their precession around the
magnetic pole, generating the well-known phenomenon of `drifting
subpulses' and hence the double cone structure. In section (6) the
drift model is applied to the pulsar PSR0943+10 and precise fits to
both the B-mode and the Q-mode are found. In section (7) the global
nature of the pulsar phenomenon is stressed: azimuthal interactions
between critical regions of the outer magnetosphere are shown to
explain the observed dependency of conal formation on the angle of
inclination. Finally, in section (8), the physical requirements of the
underlying feedback system are discussed in the light of current
physical ideas.
  
The sketched model which emerges from this analysis contains many
elements of the RS model, but on a scale which involves the entire
magnetosphere: particles (presumed here to be electrons, although the
system is sign-reversible) are accelerated from low Lorentz factors
close to the polar cap to achieve high $\gamma$ near the outer gap,
which stretches from the corotating dead zone of the magnetosphere to
the light cylinder. Pair creation occurs in these regions, although
certainly not in the profusion of the pair cascades supposed in
Crab-like pulsars (Cheng et al, 1986, Romani $\&$ Yadiaroglu,
1995). Rather the production is likely to be intermittent, weak (ie of
low multiplicity with low Lorentz factors) and azimuthally-dependent,
with most of the particles produced forming a wind beyond the light
cylinder, but with a small but essential fraction of the positrons
returning to the surface, accelerated and funnelled by the
increasingly tight bundle of magnetic field-lines above the poles
(Michel 1992).

Unlike the case of fast pulsars (Cheng et al, 1986), the downward flow
of particles is inadequate to screen the polar potential.  Shortly
before reaching the surface, the positrons emit sufficiently energetic
radiation to cause a bunched avalanche of pairs which bombard the
surface with coherent radiation. This is then reflected and
contributes to the core component of the pulsar profile. Residual
electrons formed by this process (maybe augmented by electrons emitted
from the heated spot on the surface) are then accelerated back to the
outer gap. This process is then repeated, and is reinforced if the
combined and equal drifts of the electrons and positrons around the
pole maintain the hot spots at locations which are either fixed in
azimuth or drift slowly backwards or forwards. The outflowing
electrons are somehow stratified by the inflowing layers of pairs, and
radiate curvature radiation parallel to the field lines and at about
200 km (for 1GHz) above the surface. The holistic nature of the model
means that the radio emission can simply be seen as an image of, and
as driven by, the activities of the outer gap.

The emission, particle flow and pair-creation processes are not worked
out in detail here, and in many cases have been cannibalised from
existing models (especially Mestel et al (1985), Michel (1992),
Shibata (1994), HAa,b, HM, HMZ, Hirotani $\&$ Shibata 1999, 2001
(HSa,b)). Some features lack, as yet, a proper theoretical
investigation: for example, the mechanism and level of weak low-energy
pair production in outergaps has never been explored, since hitherto
it has been assumed that outergaps only play a significant role in the
emission of fast, young pulsars. On the other hand, the recent insight
(see HAa,b, HM, HMZ, above) that in older pulsars, whose pair
production at the polar cap relies on Inverse Compton Scattering,
there will be a residual potential to accelerate particles up towards
the outergap (and by implication could accelerate returning particles
of an opposite charge) gives support to a fundamental aspect the
model.

However we strongly stress througout the paper that the direction and
purpose here is to establish the broad characteristics of a workable
model, based on observational rather than theoretical grounds. In
doing so we clarify not only what features such a model needs, but
also what is not needed: there is no polar gap, no pair-creation in
the outflow before the outer gap is reached, no self-stratification of
the outflow to generate the coherence of radio emission, and no
multipoles.

\section{Integrated profiles}
From the early days of pulsar research it has been known that despite
the complexity of individual subpulse behaviour the integrated
profiles of pulsars remain remarkably stable (Taylor $\&$ Huguenin
1971). The profiles themselves adopt many forms, and despite many
years of patient research (Rankin 1983, Lyne $\&$ Manchester 1988, RI,
RII, Rankin 1993b, Gil $\&$ Krawczyk 1996) no consistent morphology
has yet found universal acceptance. Although alternative
interpretations can be argued (e.g. the `patchy' models of Lyne $\&$
Manchester 1988, Han $\&$ Manchester 2001, Smith 2003), we will here
adopt the analyses of RI, RII, Gil et al (1993) and Kramer et al
(1994), which suggest a picture which is strikingly simple: at any
given frequency, any pulsar profile can be represented as a cut across
a notional emission envelope made up of a central core plus two
concentric cones. The core is centred on the magnetic axis above the
pulsar magnetic pole, and the cones surround it one or two hundred
kilometres above the polar cap.  Sometimes one or other of the cones
is missing.  As the frequency decreases the cones widen according to a
dipole geometry (radius-to-frequency mapping), so that the conical
structure is `tied' to the field-lines. We propose to explore the
double cone model in order to see what geometric - and hence physical
- consequences flow from it.

In Rankin's work (RI, RII) much depends on the accuracy of the
inferred values for $\alpha$, the angle between the pulsar's rotation
and magnetic axes. These are calibrated using the measured angular
width of the core component in perpendicularly rotating pulsars. This
method implicitly assumes that the core component always spans the
same field-lines, that at {\em all} alignments the core component
appears to be radially emitted from the surface in the region of open
field-lines between the boundaries of the closed `dead' corotating
zone. These assumptions have the obvious weakness, conceded by Rankin
in RI, that no physical model has hitherto been suggested with these
properties. Yet a possible explanation for this (eg Michel 1992)
emerges later from the analysis here in terms of reflected emission
from downfalling particles. However, we stress from the start that our
purpose is not to impose such a model, but merely to point out that
the geometry will support it.

The aim of the detailed analysis in RI and RII was to disentangle the
geometric coincidence created by our line of sight from the underlying
intrinsic geometry. Using the established values of $\alpha$ to infer
the geometry of the inner and outer cones, it was found that, when
present, the cones always subtend the same angular radii to the
magnetic axis, dependent only on the period P and frequency.  In a
study of some 150 pulsars Rankin (RII) concluded that the outer and
inner radii appropriate for 1GHz are
\eq
\rho_{outer}=5.75^oP^{-0.50}
\lab{1}
\en
and
\eq
\lab{2}
\rho_{inner}=4.33^oP^{-0.52}
\en
The symmetry of the conal positions about the centre of the profile is
all the more surprising since the intensities of the peaks are often
highly asymmetric (e.g. PSR1133+17 (Nowakowski 1996)).  Some pulsars
possess both cones (type M), others merely one (type T), and those
with one cone may take either the inner or the outer radius. Type
$S_t$, a young population generally exhibiting a single core
component, sometimes reveal a conal structure at high frequencies
(RII).  It is remarkable that, whatever the configuration,
intermediate values between (1) and (2) are rarely found.
Furthermore, most pulsars seem to have no more than two cones
(although four or five relatively fast pulsars may have cones
displaced onto wider cones, but in the same ratio (Mitra $\&$
Deshpande, 1999, see also Section 5.4)).

The double-cone results were apparently confirmed (with smaller
samples) by Gil et al (1993) at 1.4GHz and 10.55GHz, and Kramer et al
(1994) at 1.4GHz, 4.75GHz and 10.55GHz, yielding the expected smaller
opening angles consistent with the radius-to-frequency mapping. None
of these studies included millisecond pulsars, whose properties will
also not be considered here, in a deliberate attempt to compare only
pulsars with similar magnetic field strengths.

\begin{figure}
$$\vbox{
\psfig{figure=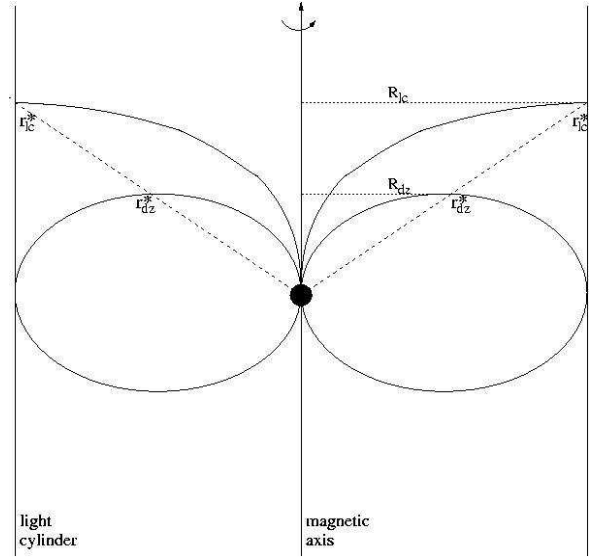,width=7.7truecm,angle=0}
}$$
\caption{The geometry of the `mirror' points in a plane projected onto
 the rotation axis. The null line intersects the dead zone and the light
 cylinder at $r_{dz}^{*}$ and $r_{lc}^{*}$ respectively.}
\label{Fig1}
\end{figure}

In interpreting her results, Rankin (RII) notes that the inverse root
dependence on period supports a dipole geometry: the radius of the
polar cap, defined by the last field-line to close within the
light-cylinder, scales with $P^{-0.5}$, and she therefore suggests
that the two cones are both formed tangential to this same field-line
at differing heights, with the emission of the inner cone lying below
that of the outer cone. Following the logic of the dipole geometry,
the outer cone at 1 GHz, whose angular radius is given in (1), is then
formed at a period-independent height of 220km above the surface, and
the inner cone at 110km.

This interpretation is difficult to reconcile with the
frequency-to-radius mapping concept, whereby greater heights
correspond to lower frequency emission, and theoretical efforts have
been made to resolve this. Petrova (2000), Petrova $\&$ Lyubarskii
(2000) and Qiao et al (2000) have suggested bimodal propagation models
for the inner magnetosphere. However these models have been devised to
explain the conal profiles, and it is as yet hard to see how they
operate at the more fundamental subpulse level without implying that
the inner and outer cones have identical subpulse behaviour (something
which is rarely, if ever, observed).

In this paper we adopt an alternative interpretation: that the inner
cone is also formed at the same height as the outer cone, but on a
field-line closer to the magnetic axis. This suggestion is not new,
and can be made consistent with the RS model by including surface
multipole components of a specific form (Gil et al 1993, Gil $\&$
Sendyk 2000, Asseo $\&$ Khechinasvili 2002). But it would also enable
the observational conclusions of Kijak $\&$ Gil (1997) and Kijak
(2001) that emission heights decrease with decreasing period to be
reinterpreted as emission cones moving closer to the magnetic axis.

Clearly identifying the true fieldlines on which emission occurs is a
formidable observational task (Mitra $\&$ Rankin 2001, Kijak $\&$ Gil
2003). But in the present context, where we are pursuing the
hypothesis of a purely dipolar field, any suggestion that different
cones are located on different fieldlines is highly radical, since it
would appear to imply that special field-lines are somehow chosen near
the surface by the upward-moving particles - and the question arises
how the particles can `know' which field-lines to select! The dilemma
can only be resolved by abandoning the concept, long-held by
theoreticians, that the structure of the magnetosphere is determined,
indeed driven, by conditions near the polar cap. The step we take here
is to reverse this logic, and argue that events in the outer
magnetosphere are the true `engine' of the pulsar system, and that the
observed polar region events are passive reflectors of these.

In their seminal work of 1969, Goldreich $\&$ Julian(GJ) divided the
polar cap into a central region of outflowing (negative) particles and
an outer annulus within which the current returned to the star. The
outer boundary was defined, as in Rankin's work, by the last closed
field line.  The inner boundary was defined by the field-line which
intersects the light cylinder at the same point as the so-called null
surface. The null surface is to first approximation the locus of the
points on which the magnetic field is perpendicular to the rotation
axis (see Mestel 1999, p536), and is physically significant since it
is the surface separating regions of net positive charge from those of
net negative charge on which the net charge density must be zero in a
steady system. In an aligned pulsar the central region has negative
charge density and the annulus positive. This polarity is reversed in
the the counter-aligned model of RS, where the polar cap does not
extend beyond the inner boundary.

The importance of the null surface was first realised by Holloway
(1973), who pointed out that a charge-separated particle flow could
never smoothly cross it, and that therefore an `outer gap' of intense
electric field would form, extending from the corotating dead zone to
the light cylinder surface. In energetic young pulsars this gap is
thought to be the site of gamma radiation (Cheng et al., 1986). In an
axisymmetric system the null surface forms a cone of half-angle
$\arctan{\sqrt{2}}\simeq{54.7^o}$, which intersects the light cylinder
at an altitude of
\eq
\lab{3}
r^{*}_{lc}={(\frac{3}{2})}^{\frac{1}{2}}R_{lc}={(\frac{3}{2})}^{\frac{1}{2}}\frac{cP}{2\pi}
\en
and the dead zone at 
\eq
\lab{4}
r^{*}_{dz}=\frac{2}{3}R_{lc}=\frac{cP}{3\pi}
\en
where $R_{lc}=\frac{cP}{2\pi}$ is the light-cylinder radius. The
field-lines linked to these extrema enter the neutron star surface at
angles which are in the period-independent ratio of
\eq
\lab{5}
{(\frac{r^{*}_{dz}}{r^{*}_{lc}})}^\frac{1}{2}=(\frac{2}{3})^{\frac{3}{4}}=0.74
\en
(inner to outer). Through dipole scaling this ratio would then apply
at any given height above the surface.

\begin{figure}
$$\vbox{
\psfig{figure=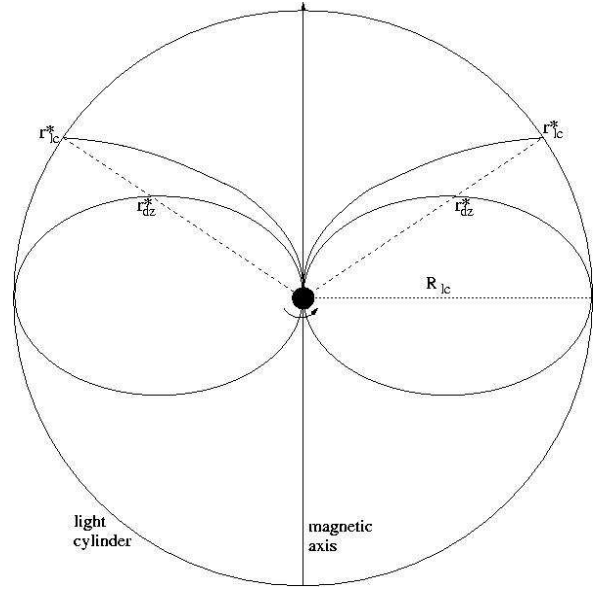,width=7.7truecm,angle=0}
}$$
\caption{The geometry of the `mirror' points of a near-perpendicular
 rotator in the plane of the observer`s line of sight. The null line
 intersects the dead zone at $r_{dz}^{*}$ as before, and the light
 cylinder at a slightly reduced $r_{lc}^{*}$.}
\label{Fig 2}
\end{figure}

Of course real pulsars are inclined rotators and in their outer
magnetospheres are far from axisymmetric and far from dipolar. Yet our
line of sight passes over the magnetic pole with a low `impact' angle
in a plane where the magnetic field close to the surface is dipolar
and hence axisymmetric about the pole to a first approximation. In
this plane the null lines (along which the magnetic field is
perpendicular the rotation axis) remain unchanged from those shown in
Fig 1 for a precise dipole, and will make the same intersection with
the dead zone at $r^{*}_{dz}$. However the light cylinder intersection
will shift by an extent dependent on the angle of inclination. Fig 2
shows the most extreme case when the the magnetic and rotation axes
are near-perpendicular, but still using a dipolar structure. The
critical field-line for the inner cone has moved slightly outwards and
the ratio of 0.74 has changed to 0.82.

In the more realistic Deutsch (1959) vacuum solution the field-lines
near the light cylinder are swept back and would give an asymmetric
and curved null lines at large inclinations. Furthermore the limiting
field-lines defining the boundary of the dead zone will be perturbed,
quite possibly in an asymmetric fashion.  But we stress that for most
pulsars these effects may only be of second order, and the ratio
$(\frac{r^{*}_{dz}}{r^{*}_{lc}})^{\frac{1}{2}}$ may not stray far from
the 0.74 to 0.82 range, especially since statistically the angles of
inclination for pulsars cluster around $35^{o}$ and are only
exceptionally found at high inclinations (see Figs 2 and 4 of RI, and
Section 7 of this paper).

Taking the observed ratio of the inner to outer cone radii at 1GHz
from (1) and (2) we obtain
\eq
\lab{6}
\frac{\rho_{inner}}{\rho_{outer}}=0.75
\en
The ratio is slightly higher at higher frequencies: 0.77 at 1.4GHz
(Gil et al 1993), 0.78 at 4.75GHz (Kramer et al 1997). These ratios
are consistent with the range indicated from the theoretical analysis
above and suggest that we might identify the critical lines of GJ with
the defining field-lines of the emission cones. This, in turn, points
to an intimate relation between locations on the null surface - whose
precise location may vary from pulsar to pulsar - and the pattern of
emission above the polar cap. Somehow the outer magnetosphere and the
polar regions mirror one another.  This is an idea not present in
polar gap models or space-charge-limited models for `slow` pulsars (eg
RS, HAa,b, HM, HMZ), yet it immediately suggests an evolutionary link
to the outergap models for fast pulsars (HSa,b). Thus we may suppose
that, although now weak, outergap pair production may still play a
crucial role in the radio emission process.

\section{Conal type and the critical field-lines} 
The principal result of RII is that the profile types ($S_t$, M, T)
show a dependence on period, yet no dependence on surface-related
parameters such as $B_s$, the surface magnetic field strength, or the
acceleration parameter $\frac{B_s}{P^2}$.  In general, inner cones are
found to form at shorter periods while outer cones form at longer
periods, with an overlap range where both cones are present.  If we
assume that the outer cones and inner cones are formed by the same
physical process at either null-line intersection, then this result
can be reproduced by a simple geometrical argument. All that is needed
is the assertion that a cone only forms when its corresponding
intersection point lies within a fixed, period-independent, range of
altitudes.

The simplest way to see this is to consider two pulsars, one of period
$P_{1}$ and a faster one of period
$P_{2}=(\frac{2}{3})^{\frac{3}{2}}P_{1}$. Then from (3) and (4) the
outer cone of the slower pulsar is formed on exactly the same
field-line as the inner cone of the faster. And since the opening
angle of the null surface is independent of period, the magnetic field
strength and the distance of the intersection from the surface remain
the same, although now the intersection is with the light-cylinder of
the faster pulsar rather than the dead zone of the slower pulsar.
Thus whether a particular field-line appears as an inner or outer cone
is determined solely by the pulsar period. This is completely
independent of whatever physical criteria must be met in order to form
a cone.

With this insight it is possible to use the observed period dependency
of the various profile types to determine the range of periods within
which emission cones are created.  Rewriting (5), the distances,
$r^{*}_{lc}$ and $r^{*}_{dz}$, of the intersection points from the
star (which are proportional to $R_{lc}$, and hence to the period) are
in the ratio
\eq
\lab{7}
\frac{r^{*}_{lc}}{r^{*}_{dz}}=(\frac{3}{2})^\frac{3}{2}\simeq{1.84}
\en
Let us then suppose that a cone only forms when the intersection
points fall within $r^{*}_{min}$ and $r^{*}_{max}$. For $r_{dz}^{*}$
to be in this range, and hence for an outer cone to form, the pulsar's
period must, from (4), satisfy
\eq
\lab{8}
3\pi\frac{r_{min}^{*}}{c}<P<3\pi\frac{r^{*}_{max}}{c}
\en
Similarly, the period range for inner cone formation is, from (3),
\eq
\lab{9}
2({\frac{2}{3}})^{\frac{1}{2}}\pi\frac{r{*}_{min}}{c}<P<
2({\frac{2}{3}})^{\frac{1}{2}}\pi\frac{r{*}_{max}}{c}
\en
The first range maps onto the second by the factor (7). As is shown in
Fig 1, long-period pulsars with wide light-cylinders will be found to
have only an outer cone because the light-cylinder intersection lies
above $r^{*}_{max}$, and fast pulsars only an inner cone bcause the
deadzone intersection is below $r^{*}_{min}$, with pulsars with an
intermediate range of periods having two cones. This is exactly as
reported in RII.

In RII Rankin finds that type $S_t$ and type T with only an inner cone
have a mean period around 0.5s, type M with two cones have a mean
period of 0.86s, and types T and D(double) with only an outer cone
have a mean period of about 1.25s. If we assume that cones are only
formed when the null-line intersection point lies between
$r^{*}_{min}=20,000km$ and $r^{*}_{max}=70,000km$, then the range for
inner cone formation is 0.32s to 1.15s, and consequently 0.6s to 2.15s
for outer cones. This implies a double cone (M) range from 0.6s to
1.15s. These figures successfully reproduce the observed mean values,
although the scatter around the mean is fuzzy (see tables in Rankin
1993b), and other hitherto unconsidered factors may be involved. Later
in this paper it is suggested that the degree of electric screening
near the magnetic pole may be such a factor, resulting, for example,
in core-dominant type $S_t$ pulsars having a significantly lower range
than the other categories.

\begin{figure}
$$\vbox{
\psfig{figure=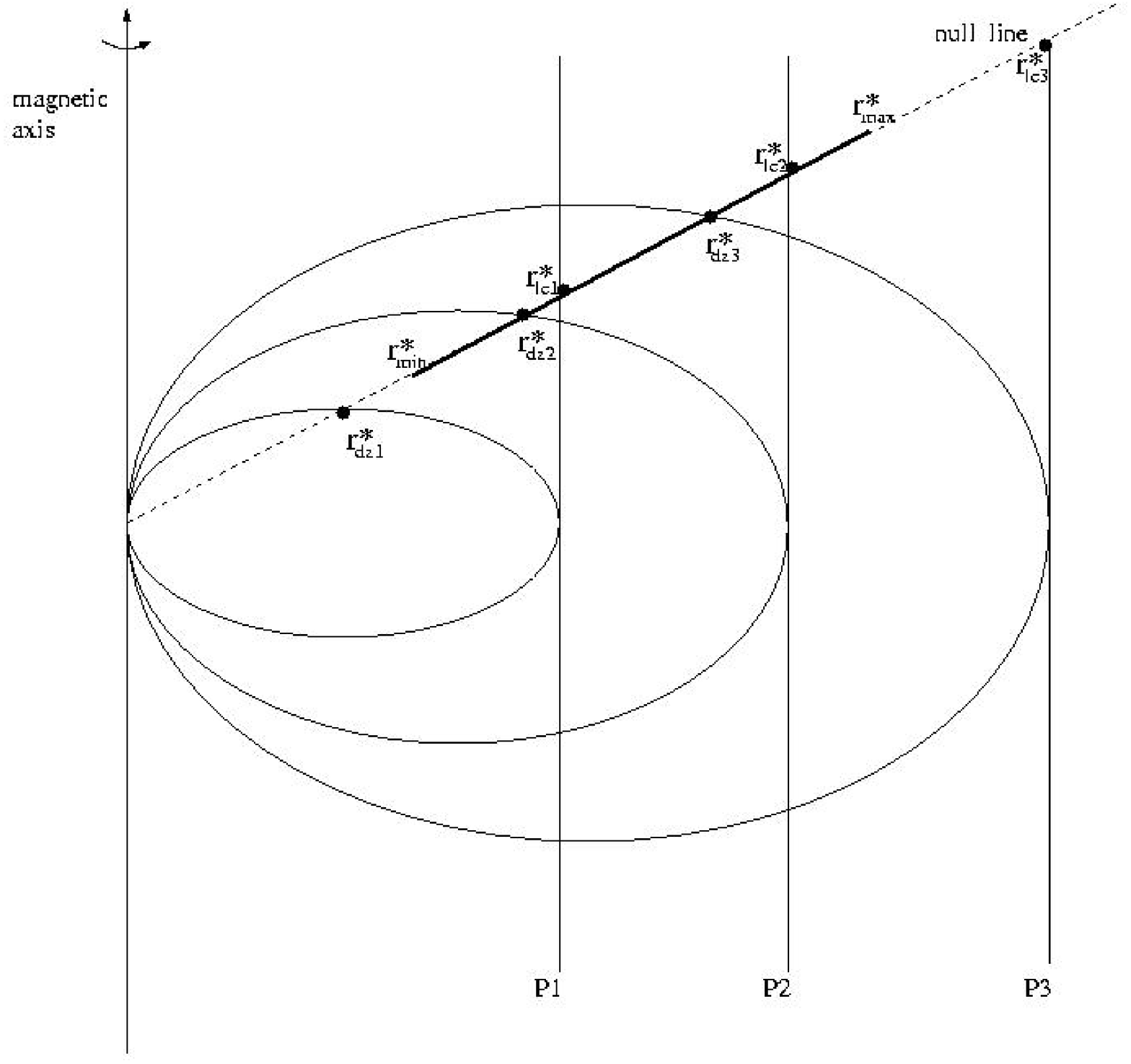,width=8.4truecm,angle=360}
}$$
\caption{the vertical line (P1,P2,P3) show different possible positions
 for the light cylinder and their intersections with the null line
 ($r^{*}_{lc1},r^{*}_{lc2},r^{*}_{lc3}$). Each gives rise to a dead
 zone which intersects the null line at
 ($r^{*}_{dz1},r^{*}_{dz2},r^{*}_{dz3}$) respectively. Only
 $r^{*}_{dz2}, r^{*}_{lc1},r^{*}_{dz3},r^{*}_{lc2}$ lie within the
 fixed range of heights indicated by $r^{*}_{min}$ and $r^{*}_{max}$
 and therefore become mirror points}
\label{Fig3}
\end{figure}

Note that it is the distance from the star which determines the
appearance or non-appearance of a cone, and not, for example, the
ambient magnetic field strength. This would depend on $B_s$, a factor
which Rankin finds not to be statistically significant.  Assuming a
surface field of $B_s=10^{12}G$, the fixed range in radial distance
translates into a poloidal magnetic field range of around
$B_{min}=100G$ to $B_{max}=4000G$, a factor of
$(3.5)^3\simeq{40}$. Actual estimates of $B_s$ vary by a factor
comparable to this, and a criterion based only on field strength at
the intersections could not reproduce the precision of the observed
range without showing dependence on $B_s$.  We are left with a
criterion based on a tight range in intersection altitudes, or,
equivalently, in star to null line communication times from 0.07s to
0.23s.

We stress that as yet no particular model for the creation of emission
zones has been imposed.  The whole purpose of this paper is to use the
observations to define and constrain the emission model. What has been
shown so far is that the observed dependence of profile morphology on
period (and period alone) can be explained by a specifying a fixed and
relatively narrow range of altitudes within which the `mirror' site
can fall.  Thus the problem of morphology has been decoupled from the
problem of devising an emission model. Observations seem to be forcing
us to consider pulsar radio emission as a global pan-magnetosphere
phenomenon, involving intercommunication between the polar cap and the
outer gap.  The required features of an emission model based on this
interpretation are discussed in the next section.

\section{The formation of emission nodes}
In the previous sections it is argued that the observations of radio
profiles support the existence of a feedback system betwen the star
and the mirror points.  This implies a flow of charged particles in
{\em both} directions, guided by the magnetic fieldlines. Unless the
electric fields parallel to the magnetic fields are universally
screened (we discuss in Section 7 recent work by Harding and coworkers
(HM,HMZ), who indeed find that in older pulsars this potential cannot
be fully screened) it inevitably follows that the streams of charged
particles will undergo an azimuthal drift, irrespective of how they
are produced at either end. The sense of the drift will be the same
for particles of either sign, and independent of whether they are
moving outwards or inwards. It is also independent of whether we are
dealing with a pulsar with rotation axis and magnetic moment in the
same sense (as in the `inner accelerator' models) or with an
antipulsar (as in RS), where the axes are in opposing directions.

The picture we adopt is as follows: electrons are gradually
accelerated outward on the critical mirror fieldlines from close to
the surface to the the mirror points on the null line, where they
initiate pair-production in the outergap. This production will not and
cannot be prolific since the interacting photons will have low energy
and will fail to initiate the massive pair cascades found in younger
pulsars (Cheng et al 1986, HSa,b).  Nevertheless sufficient low-energy
positrons are then supposed to be available, either through
pair-production or through interaction with another magnetospheric
region (as envisaged in the previous section), or both, to cause a
backflow of positive charge to the surface. The total current flow
will be dominated by electrons, since in a steady state the net charge
density must always be negative and close to the charge-separated GJ
value. The drift will cause the positrons to return to the surface on
a different fieldline from which the electrons came (see Fig 4), and
they will be accelerated towards a different azimuthal location on the
polar cap. Close to the surface the positron energies will be
sufficient to trigger a pair-production `avalanche' (as described in
Michel 1992), which will cause electrons, either from the surface or
trailing the avalanche (or both) to be accelerated upward into the
magnetosphere. These will, in turn, continue to drift on their passage
to the mirror regions. If theory can demonstrate that no particles
from the surface are needed or even possible in this scenario, the
entire model becomes sign-reversible and can apply to pulsars and
antipulsars alike.

The central proposal here is that in a quasi-steady state the
avalanches above the surface will be confined to discrete `emission'
nodes, arranged so that the flow forms a continuous yet finite stream
of particles linking all the nodes (Fig 4). It is presumed that by
remaining separated they avoid mutual interference and maintain
conditions for pair-creation and the observed coherent radiation. In
this section we describe and quantify the process by which these nodes
are formed, and in the next section we consider how observations of
subpulse drift can be interpreted as a movement of these nodes and be
used as diagnostics of the conditions in a pulsar's magnetosphere.

In the previous section it is argued that observations suggest that a
high level of interconnectivity between the poles is a prerequisite
for pulsar emission. Thus we may expect the particles created at the
mirror points will feed both poles and ensure that particle drift at
one pole is coordinated with that at the other. This supports a view
that the mirror points are far from being the passive reflectors of
events close to the surface, but rather drive the entire drifting
phenomenon by means of null line interactions between the cones, and
possibly through inter-pole links. Here we analyse an axisymmetric
system at a single pole, although the ultimate intention is to apply
the resulting estimates in the context of an inclined interactive
magnetosphere.

The particle drift rate is estimated by an argument analogous to that
of RS, but applied to an entire axisymmetric magnetosphere rather
than confined to a polar cap. Inevitably the estimates will be less
precise, but nonetheless useful.  We consider first a configuration
with an outer cone only (i.e. with no inner cone screening). If there
is a potential drop of ${\Delta}V$ along the central magnetic axis
from the surface to a height at the level of the dead zone
intersection (see Fig 1), then the typical potential difference
available over the cylindrical radius of $R_{dz}$ between the axis
and critical fieldlines close to the deadzone is $-{\Delta}V$, since
there is a zero potential drop along field-lines bounding the dead
zone and across the surface of the polar cap. Then taking the typical
scale length of the electric field in the region of the dead zone as
$R_{dz}$ we obtain $\frac{{\Delta}V}{R_{dz}}$ as a measure of the
mean non-corotational electric field perpendicular to the magnetic
field $B_{dz}$ at a cylindrical radius of $R_{dz}$. Using this we
can estimate the mean magnetospheric drift rate relative to the
corotating frame, and in the opposite sense to the star's rotation,
as

\begin{eqnarray}
\mid{(\Omega_{D})^{'}_{out}}\mid&=&\frac{1}{R}\frac{\mid{\bf \Delta{E}XB}\mid}{B^2}c\nonumber\\
&\approx&{\frac{c{\Delta}V}{B_{dz}R_{dz}^2}}=\frac{2\pi}{P}\frac{{\Delta}V}{({\Delta}
V_{max})_{out}}
\end{eqnarray}
where ${\bf \Delta{E}}$ is the variation in the electric field from
the corotational value, and
\eq
\lab{11}
({\Delta}V_{max})_{out}=\frac{2\Phi_{out}}{cP}\approx{\frac{{2\pi}B_{dz}R_{dz}^{2}}{cP}}
\en
is the maximum potential available in the vacuum case, and
$\Phi_{out}$ is the magnetic flux through the polar cap extending to
its dead zone boundary (Sturrock 1971). Equation (10) gives zero net
drift in the inertial frame when ${\Delta}V=({\Delta}V_{max})_{out}$,
and corotation if ${\Delta}V=0$. Note that $(\Omega_D)^{'}_{out}$ is
formally negative. The net particle drift rate in the inertial frame
($(\Omega_D)_{out}=\frac{2\pi}{P} + (\Omega_D)^{'}_{out}$) is thus
proportional to the level of screening within the magnetosphere, a
quantity which may well be time-dependent. Henceforth, primed
quantities will indicate that these are measured in the corotating
frame.

An equivalent estimate can be made for the inner cone drift, although
this will be more approximate since the field-lines will be swept back
with a significant toroidal component, and particles are likely to
have sufficient energy to leave the field-lines as they cross the
light cylinder (Mestel et al, 1985). Assuming the potential difference
of $\Delta{V}$ along the central axis now extends to a height at the
level of the light cylinder/ null line intersection yields a potential
difference $-{\Delta}V$ between the magnetic axis and the light
cylinder at $R_{lc}$. This effectively assumes a zero potential drop
along the null line between $r^{*}_{dz}$ and $r^{*}_{lc}$.  In a more
realistic model (i.e. that of Mestel et al 1985 or Shibata 1990) this
potential drop is not zero and drives the closure of the current loop
outside the light cylinder.  Using now $R_{lc}$ as the scale length,
this will produce the parallel formula to (10), namely
\eq
\lab{12}
\mid{(\Omega_{D})^{'}_{in}}\mid\approx{
\frac{c{\Delta}V}{B_{lc}R_{lc}^2}}=\frac{2\pi}{P}\frac{{\Delta}V}{({\Delta}
V_{max})_{in}}
\en
with $({\Delta}V_{max})_{in}$ calculated from the magnetic flux
$\Phi_{in}$ through the inner cone:
\eq
\lab{13}
({\Delta}V_{max})_{in}=\frac{2\Phi_{in}}{cP}\approx{\frac
{{\pi}B_{lc}R_{lc}^{2}}{cP}}
\en
$\Phi_{in}$ is less than the full polar cap flux $\Phi_{out}$ by a
factor of $\frac{r^{*}_{dz}}{r^{*}_{lc}}$, which from (5) gives
\eq
\lab{14}
(\Omega_{D})^{'}_{in}=\frac{r^{*}_{lc}}{r^{*}_{dz}}(\Omega_{D})_{out}
=(\frac{3}{2})^{\frac{3}{2}}(\Omega_{D})^{'}_{out}=1.84(\Omega_D)^{'}_{out}
\en
Thus in the simplified magnetospheric model used here the particle
drift rate in the inner cone is nearly twice as fast as that in the
outer cone, if either is present alone.

Having established the parameters of the particle drift, we are now in
a position to explore how many emission nodes are created. The
inflow/outflow process in the outer cone will, in the corotating
frame, circulate once around the cap in a time
$\frac{2\pi}{(\Omega_{D})^{'}}$. It is conjectured that the
circulation sets up regular cycle, whereby the flow `visits' $p'$
nodes within $q'$ turns and then repeats itself continuously (Fig
4). It follows that in each turn the flow makes $\frac{p'}{q'}$
descents to the cap, where
\begin{eqnarray}
\frac{p'}{q'}\approx\frac{2\pi}{\mid{(\Omega_{D})^{'}_{out}}\mid}\frac{c}{2r^{*}_{dz}}&=&
\frac{3\pi}{2}\frac{2\pi}{P}\frac{1}{\mid{(\Omega_D)^{'}_{out}}\mid}\nonumber\\
&=&\frac{3\pi}{2}\frac{(\Delta{V_{max}})_{out}}{\Delta{V}}
\end{eqnarray}
For the inner cone the equivalent calculation is
\begin{eqnarray}
\frac{p'}{q'}\approx\frac{2\pi}{\mid{(\Omega_{D})^{'}_{in}}\mid}\frac{c}{2r^{*}_{lc}}&=&
(\frac{2}{3})^\frac{1}{2}\pi\frac{2\pi}{P}\frac{1}{\mid{(\Omega_D)^{'}_{in}}\mid}\nonumber\\
&=&\frac{4\pi}{9}\frac{(\Delta{V_{max}})_{out}}{\Delta{V}}
\end{eqnarray}
The relations between n and $(\Omega_D)^{'}$ in the first lines of
(15) and (16) are purely geometric relations, showing how
$\frac{p'}{q'}$ depends on the drift rotation rate, the distance to
the nodes from the mirror points and the period of the pulsar in the
corotating frame. As will be seen in the next section, they enable us
to set lower limits on $\frac{p'}{q'}$ in both cones, since in neither
cone can $(\Omega_D)^{'}$ exceed the pulsar's own rotation speed and
counter-rotate (Ruderman, 1976).

The second lines of equations (15) and (16) express how
$\frac{p'}{q'}$ measures the extent to which the optimum potential has
been screened: the lower ${\Delta}V$ falls the larger $\frac{p'}{q'}$
will become.  To compute $\frac{p'}{q'}$ we need estimates of
$({\Delta}V_{max})_{out}$ and ${\Delta}V$. $({\Delta}V_{max})_{out}$
is obtained from (11):
\eq
\label{17}
({\Delta}V_{max})_{out}\approx{1.35}\times\frac{B_{12}}{P^2}10^{13}eV
\en
where $B_{12}$ is the surface magnetic field in units of $10^{12}G$.
For ${\Delta}V$ we can take a normalising value of, say,
${\Delta}V=5\times{10^{12}}$, which corresponds to the potential
needed to accelerate electrons (positrons) to approximately
$\gamma=9\times{10^6}$ at which radiated gamma-rays may trigger
pair-creation. However it must be stressed that the intention of the
model here is not to impose a value for $\Delta{V}$ or $\gamma$, but
to find a way to {\em measure} them. Inserting (17) and the estimate
for $\Delta{V}$ we obtain from (15) and (16)
\eq
\label{18}
\frac{p'}{q'}\approx{13}\frac{B_{12}}{P^2}(\frac{5\times10^{12}}{{\Delta}V})
\en
for the outer cone, and
\eq
\lab{19}
\frac{p'}{q'}\approx{4}\frac{B_{12}}{P^2}(\frac{5\times10^{12}}{{\Delta}V})
\en
for the inner cone.

We can now return to our result of Section (3): that emission cones
tend only to form when their mirror points lie between two fixed,
period-independent altitudes. With the model here, the upper limit may
be interpreted as a maximum distance of the mirror points above the
emission nodes at which the required coherence of the emission nodes
can be maintained close to the surface - a high degree of angular
precision must be required to confine the node to its particular
(possibly drifting) azimuth and prevent its `smearing'. At the other
extreme, a too-dense clustering of nodes may cause interference
between the electron and positron flows of adjacent nodes, preventing
a steady flow between the surface and the mirror points.

\begin{figure}
$$\vbox{
\psfig{figure=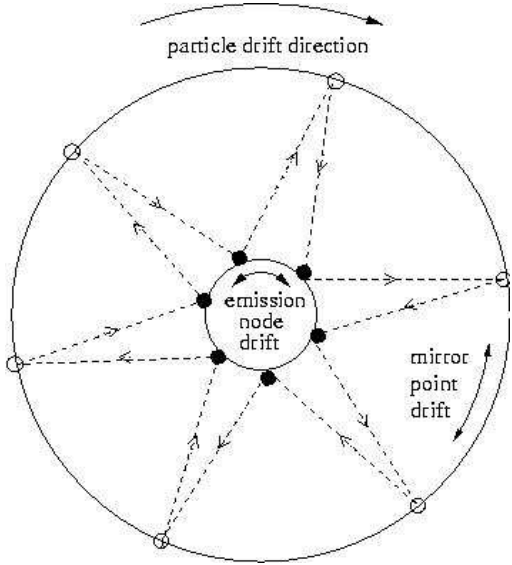,width=6.7truecm,angle=0}
}$$
\caption{A view of the particle drift looking down onto the
 magnetic pole in the corotating frame. The motion of particles over
 distances of more than 20,000km to and from the `mirror' points on
 the null line creates emission nodes immediately above the polar
 cap. The nodes, and their corresponding mirror points, may be
 stationary, or drift in either direction. The pattern shown
 corresponds to the simplest mode (i.e. all the nodes are created in a
 single turn).}
\label{Fig4}
\end{figure}

\section{Drifting subpulses}
\subsection{Introduction}
The emission of many pulsars exhibits a systematic modulation known as
drifting subpulses, whereby a succession of subpulses, identified here
as the emission of adjacent nodes, appear in the pulse window and each
gradually `drifts' in the same sense across the window to the position
of his neighbour within a timespan of $P_3$ rotation periods. The
drift may be fast ($\approx$ 2 periods) or slow ($>10$ periods), and
the pattern repeats itself, often for long stretches, before sometimes
switching to a different `mode' of emission.

In interpreting observations of drifting subpulses, the model
suggested here shares with RS the picture of emission columns
circulating around the magnetic pole. However the underlying flow
which produces these columns, illustrated in Fig 4, is radically
different and has greater flexibility. A steady system of stable
emission nodes is created and maintained near the surface by a
continuous, or quasi-continuous, flow of particles to and from the
mirror points. Depending on the rate of particle drift, which in the
corotating frame can have only one sense, the nodes may be stationary,
or precess positively or negatively around the magnetic pole. An equal
number of pair-producing regions on the outer gap, either at the dead
zone boundary ($r^{*}_{dz}$) or at the light cylinder ($r^{*}_{lc}$),
will precess in tandem (Fig 4). In theory a very wide range of
emission systems can be generated in this manner, some highly complex
and chaotic.  However the `mode' selected by the pulsar will reflect
the value of $\frac{\Delta{V}}{\Delta{V_{max}}}$.

There is a fascinating everyday analogy to the process suggested
here. If a camcorder linked to a television set is pointed at the
screen in a completely darkened room, and if an initial instantaneous
flash (such as the striking of a match) occurs between the screen and
the camcorder, then the image of the flash appears on the screen and
feeds back into the camcorder. The image on the screen will then
constantly change, creating patterns which may be regular or chaotic,
depending on the angle at which the camcorder is held about its
axis. It is a classical demonstration of how complexity can arise even
in a simple feedback system: complexity of outcome does not imply
complex input.

In the previous section we estimated the particle drift-rate in the
corotating frame. This is the appropriate frame for the calculation of
the driving electric field, but the observed number and drift of the
nodes depend critically on the observer's frame of reference. For
example, if the observer were rotating at a rate less than, but in the
same sense as, the corotating drift-rate of the particles, their
apparent drift relative to the observer would be smaller than in the
corotation frame, and he would see a larger number of nodes for every
rotation in his frame. These nodes would remain in the same sequence
as in the corotating frame, but would begin to repeat before a single
turn was completed. If the observer moved counter to the drift, he
would see fewer nodes in a single turn, but on the second and later
turns the nodes would appear to overlap, and the order of the nodes
round the axis would therefore be changed. This is an important point,
since the pulsar observer in his inertial frame will always turn
counter to the sense of the particle drift in the corotating frame,
and may well observe a node pattern very different from that in the
corotating frame (as will be seen in the case of PSR0943+10, discussed
in the next section).

\subsection{Critical parameters}
We consider a distant observer's view of the drift, and relate the
underlying particle drift in his inertial frame to the number of nodes
and to the observable quantity $P_3$. We begin by adapting the
estimates in (15) and (16) to the inertial frame drift
$(\Omega_D)_{out/in}=\frac{2\pi}{P}+(\Omega_D)^{'}_{out/in}$, but
still assuming that the pulsar is aligned.  In this frame the observer
sees p nodes created in q turns by successive inflows and outflows
(where p and q are integers with no common factors) so that every q th
node is sequentially linked as the particle flow circles the
pole. Then
\eq
\lab{20}
\frac{p}{q}\approx\frac{2\pi}{\mid{(\Omega_{D})_{out}}\mid}\frac{c}{2{r}^{*}_{dz}}
\en
for the outer cone, and
\eq
\lab{21}
\frac{p}{q}\approx\frac{2\pi}{\mid{(\Omega_{D})_{in}}\mid}\frac{c}{2{r}^{*}_{lc}}
\en
for the inner cone. Note that $\frac{p}{q}$ and $\frac{p'}{q'}$ in the
notation of the previous section are equivalent parameters in the
inertial and corotating frames respectively.

Each of the particle path distances ${r}^{*}_{dz}$ and ${r}^{*}_{lc}$
defines a mean transit time ($\tau_{dz}$ or $\tau_{lc}$ in units of P,
the pulsar rotation period) from the nodes to the mirror points and
back in the outer and inner cone respectively, so that
\eq
\lab{22}
\tau_{dz}=\frac{2{r}^{*}_{dz}}{Pc}\approx\frac{2}{3\pi}=0.212
\en
for the outer cone, and for the inner cone
\eq
\lab{23}
\tau_{lc}=\frac{2{r}^{*}_{lc}}{Pc}\approx(\frac{3}{2})^{\frac{1}{2}}\frac{1}{\pi}=0.395
\en
where the estimates are those for the aligned rotator as in (3) and
(4). The inner cone value is almost double that of the outer cone,
reflecting the greater distance the flow has to travel to the light
cylinder intersection.

In practice, q will be taken as low ($<10$) and any more complex flow
will be seen as a modulation of a basic (p,q) pattern. Since the
particle drift in the inertial frame cannot exceed the rotation speed
(Ruderman 1976), we demand that
$\mid{(\Omega_D)_{out/in}}\mid<\frac{2\pi}{P}$. Hence from(20) and (21)
\eq
\lab{24}
\frac{p}{q}>\frac{1}{\tau_{dz}}\approx\frac{3\pi}{2}=4.7
\en
for the outer cone, and
\eq
\lab{25}
\frac{p}{q}>\frac{1}{\tau_{lc}}\approx(\frac{2}{3})^\frac{1}{2}\pi=2.6
\en
for the inner cone.

\subsection{Conditions for steady subpulse drift}
To turn the approximate results (20) and (21) into exact equations we
need to incorporate into $(\Omega_D)_{out/in}$ a term which takes
account of the rate at which the nodes move around the magnetic axis.
After a total time of $\hat{P}_3=p{P_3}P$ (following the notation of
RS for $\hat{P}_3$) the entire ring of emission nodes will have
presented itself to the observer. Thus in a smoothly-flowing system
(20) can be rewritten as
\eq
\lab{26}
(\Omega_{D})_{out}=\frac{2\pi}{\tau_{dz}}\frac{q}{p}+\frac{2\pi}{{\hat{P}}_3}
=\frac{2\pi}{P}\times\frac{q}{p}[\frac{1}{\tau_{dz}}+\frac{1}{q{P_3}}]
\en
In the equivalent expression for $(\Omega_{D})_{in}$, $\tau_{lc}$
replaces $\tau_{dz}$. As long as $\mid{P_3}\mid>1$, the first term in
the brackets of (26) always dominates over the second. Hence the drift
will be only a fraction of the true particle drift. In general, there
is no preference for the sign of $P_3$, and the observed drift
may appear to be in the sense of rotation or the reverse. When both
inner and outer cones are present, their nodes may conceivably have
differing drift directions.  Opposing drifts at differing longitudes
of the pulse window have been observed in PSR0540+23 (Nowakowski,
1991).

The picture implied by (26) is of a simple integrated system with a
near-continuous even flow of particles streaming from each node up to
the mirror points and back, successively visiting all of the drifting
nodes in the system cycle time (evaluated in (28) below) before
returning to meet the node from which it set out. More realistically,
the flow may be made up of `packets', possibly separated by the
node-mirror point travel time ($\approx{0.1P}secs$), but sufficiently
continuous for us to `see' each node at our sampling rate of P
seconds. Should this occur the observed emission could be separately
modulated by phase and by intensity.

In practice, observers will be able to measure $P_3$ (though not
necessarily its intrinsic sign), possibly also p (as in PSR0943+10),
and will wish to infer q and $\tau_{dz/lc}$. From the analysis so far
they will have at hand conditions (24/25), and equation (26). But
there is one further useful constraint which links the known and
unknown quantities. This essentially geometric point arises because
the magnitude of the residual angle drift, represented by the last
term in (26), must be small enough (i.e. $\mid{P_3}\mid$ large enough)
to ensure that the system has just p nodes and does not
accelerate/decelerate to a (p-1) or (p+1) system (depending on the
sign of $P_3$). In other words, for a given p, q and $\tau_{dz/lc}$
there is a critical $\mid{P_3}\mid$ below which a steady system cannot
form.

In establishing this condition we will examine in greater detail, in
equations (27) to (34), the geometric and physical features of a
near-steady conal system which underlie the result (32). The following
equations apply to either cone, but are illustrated using outer cone
parameters.

Consider particles being emitted simultaneously from all p nodes. The
flow from each node will then call at its next successive (not
necessarily adjacent) node after just $\tau_{dz}\approx{0.212}$
periods, the fundamental unit of time for the outer cone subpulse
system of the pulsar, which will in general depend on both P and
$\alpha$ in an non-aligned system. In this time the node will appear
to advance by
\eq
\lab{27}
\frac{2\pi}{\hat{P}_3}\tau_{dz}\approx\frac{4}{3pP_3}radians=\frac{76.4^o}{pP_3}
\en
The node will drift smoothly if this angle is small compared with the
angular extension of the node. The flow will then continue to the next
node, and by the time the round trip is complete, all p nodes will
have been visited in the system cycle time
\eq
\lab{28}
t_c=p\tau_{dz}\approx{0.212pP} secs
\en
and the flow will have traversed a total angular distance of
\eq
\lab{29}
2\pi{q}+\frac{2\pi}{\hat{P}_3}t_c=2\pi{q}+\frac{2\pi}{P_3}\tau_{dz}\approx{2\pi{q}+\frac{4}{3P_3}}
\en
This result can be equivalently obtained from (26) by multiplying
$(\Omega_D)_{out}$ by $t_c$.

The residual final term in (29) (which we will refer to as the cycle
drift angle) is given by
\eq
\lab{30}
\Theta_c=\frac{2\pi}{P_3}\tau_{dz}\approx{\frac{4}{3P_3}}=\frac{76.4^o}{P_3}
\en
and represents the net angle through which the system has drifted in a
single cycle. It is a useful observable parameter in determining the
characteristics of the system, and it enables us to determine how many
cycles, $N_c$, are required to rotate the entire system through a full
circle, namely
\eq
\lab{31}
N_c=\frac{2\pi}{\Theta_c}=\frac{P_3}{\tau_{dz}}\approx\frac{3\pi{P_3}}{2}=\frac{P_3}{0.212}
\en
$N_c$ must be an integer if the system is to be closed and have a
simple repeatability. Then any `streaky' features in the underlying
flow will share the $\hat{P}_3$ periodicity, the phase periodicity of
the nodes, since
\eq
\lab{32}
\hat{P}_3=N_c{t_c}
\en
For the inner cone the net drift angle, $\Theta_c$, is significantly
larger ($\frac{141^o}{P_3}$) and $N_c\approx\frac{P_3}{0.395}$.

A further significance of the cycle drift angle is that it can
determine, for a given $P_3$, how close the p-node system is to
going over into an $p+1$ or $p-1$ system. As $P_3$ reduces, the
cycle drift angle grows (see (30)), and if the system is to keep its
identity as a rotating p-node pattern, $\Theta_c$ must not exceed
the separation ($\frac{2\pi{q}}{p\pm{1}}$) between successive nodes
in the $p+1$ or $p-1$ system (depending on the sign of $P_3$). Using
(31), a minimum $P_3$ for a given p is therefore set by
\eq
\lab{33}
\mid{P_3}\mid>\tau_{dz}\frac{(p\pm{1})}{q}\approx
\frac{2(p\pm{1})}{3\pi{q}}=0.212\frac{(p\pm{1})}{q}
\en
where the positive sign applies if $P_3$ is negative, or more simply,
\eq
\lab{34}
N_c>\frac{p\pm{1}}{q}
\en
Combined with the conditions (24)/(25), (33) and its inner cone
equivalent enable useful limits on p and q to be set. It is
immediately clear that for simple systems with q=1, a high value for p
will require a high $P_3$ (i.e. a low drift-rate): for example, if
p=20 in an outer cone geometry, $P_3$ must exceed 4.  This point is
particularly relevant to PSR0943+10, discussed in the next section.

Using his estimate for (p,q) and taking a plausible value for
$\tau_{dz}$, the observer could now evaluate the drift-rate of the
particles, and hence infer the potential difference $\Delta{V}$ from
(18). But in pulsars which exhibit great regularity in their subpulse
drift it is possible to take one further step. Note that a
particularly simple class of drift states can exist if the pulsar
adopts a $P_3$ such that the value of the square bracket in (26) is an
integer or a simple fraction. This would result in a harmonic relation
between the underlying particle drift and the star`s rotation rate, a
factor which would make the flow steady, sustainable and hence
conducive to the formation of emission nodes. It may even be a
precondition for a pulsar to be observable. In an outer cone,
$\frac{1}{\tau_{dz}}$ is 4.7 in the aligned dipole case (from (22)),
and the mean value of this term will not differ greatly at
inclinations with low $\alpha$. This suggests that in regularly
drifting pulsars the value of the square brackets is 5, and that the
harmonic relation to the rotation rate would be especially clear if p
were a multiple of 5. Furthermore $P_3$ has to be positive in
nearly-aligned pulsars. From (26) and (31) the harmonic condition
implies values for $P_3$ and $\tau_{dz}$ of
\begin{eqnarray}
P_3&=&\frac{1}{5}(N_c+\frac{1}{q})\\
\tau_{dz}&=&\frac{1}{5}(1+\frac{1}{qN_c})
\end{eqnarray}
Pulsars which exhibit highly regular drifting subpulses may be
expected to have parameters which satisfy these equations.

\subsection{Mode-changing}
Many well-known pulsars maintain a steady or near-steady drift
behaviour over many periods (even over thousands of periods in the
case of PSR0943+10 discussed in the next section). These pulsars
appear to have low values of the inclination angle $\alpha$ (RI), and
hence the axisymmetric model of the previous section would seem a
reasonable approximation of their conditions. Examples are PSR0031-09
(Vivekanand $\&$ Joshi 1997, Wright $\&$ Fowler 1981b), PSR1944+17
(Deich et al. 1986), PSR2319+60 (Wright $\&$ Fowler 1981a), PSR
1918+19 (Hankins $\&$ Wolszczan 1987), and, most recently, PSR0809+74
(Lyne $\&$ Ashworth 1983, van Leeuwen et al 2002).

However each of these pulsars also has at least one alternative - and
equally stable - drift mode, so a number of steady solutions to
equation (26) must exist in a single pulsar. By inserting integral
values of $N_c$ and q into (35) and (36) an infinite series of
discrete possible stable harmonic states can be generated. Which of
these is selected by a particular pulsar - at a particular time - may
be determined by allowed values of $\Omega_D$ and $\tau_{dz}$. These
are likely, in turn, to depend on the current flow generated from the
emission nodes close to the polar cap and on the precise height of the
null line above them (which are interrelated parameters according to
the recent outergap model of HSa,b). In the outer cone, particularly
simple harmonic ratios (26) between $\Omega_{D}$ and the rotation of
the star (such as
$\frac{1}{2}$,$\frac{1}{3}$,$\frac{2}{3}$,$\frac{1}{4}$,
$\frac{3}{4}$, etc) may only be attainable if if p is a low multiple
of 5 and not a multiple of q. A mode change to another harmonic state
can occur providing $N_c$, p and q satisfy the constraints (24) and
(34). A change in $N_c$ will cause a change in $P_3$, a change in p
will cause a change in $P_2$, the observed frequency-dependent
subpulse separation, and a change in $\frac{q}{p}$ will change the
harmonic ratio (26).

Mode changes may (from (36)) also be accompanied by changes in
$\tau_{dz}$, the particle travel time between the nodes and the mirror
points and back, therefore requiring a shift in the emission region.
Studies of the mode changes in PSR0031-07 by Vivekanand $\&$ Joshi
(1997) and Wright $\&$ Fowler (1981b) suggest a progressive shrinking
of the profile from mode to mode, which may support this
interpretation. More recent work by van Leeuwen et al (2003) argues
for a similar effect in PSR0809+74. However in all the pulsars
identified above the drift repetition rate, $P_3$, although stable for
a while, shortens abruptly and successively through two or three
discrete values before returning to the highest $P_3$ in a
quasi-cyclical manner. A possible model which emerges for this class
of pulsars is one of progressively decreasing levels of screening and
build-up of potential, followed by a sudden return to high screening,
and a cycle is established.

The quasi-stability of the general chaotic systems suggests that, once
achieved, the pulsar's potential will remain in such a state, subject
only to a gradual secular change, then return to either the original
state or some `closer' alternative mode. Key to the stability of the
subpulse behaviour during these transitions is the speed with which
the nodes reconfigure themselves following a change: if the relaxation
timescale is too long, the pulsar emission may never settle down to a
steady pattern. The intermittent, but non-periodic, activity observed
in the the core regions of many pulsars (Seiradakis et al. 2000),
whatever its origin, is evidence that most magnetospheres are indeed
in a quasi-chaotic state.

\section{A Model for PSR0943+10}
\subsection{Background}
PSR0943+10, with a period of 1.1sec and $B_{12}=2$ (see DR), has very
typical pulsar parameters, but exhibits an extraordinarily precise
alternating drifting pattern (in the dominant B mode) and occasionally
switches to an apparently disordered Q mode (Suleymanova $\&$
Izvekova, 1984).  It is of great interest here partly because the
inferred angle of inclination between the magnetic and rotating axes
is just $11^o-15^o$, suggesting comparison with the aligned model used
to estimate paramters in the previous section. But it is also of
interest since Deshpande and Rankin (2000)(DR), in their recent
detailed study of this pulsar, have provided the only example so far
where N, the total number of the emission nodes, has been directly
deduced from a precise measurement of the circulation rate. The
exceptionally stable subpulse pattern with $P_{3obs}=1.867$ is shown
to have exactly N=20 drifting nodes, with the observed drift in the
same sense as the presumed ${\bf E\times{B}}$ particle drift.

An interpretation in terms of the present model is somewhat hampered
by the fact that the line of sight trajectory of this pulsar is too
oblique for observers to determine conclusively whether we are seeing
the geometry of an inner or an outer cone (DR give two alternative
good fits with $\alpha=11.6^o$ for an inner cone, and $\alpha=15.4^o$
for an outer cone). However the facts that $P_3$ is positive and that
the pulsar is almost aligned suggest (from the discussion in Section
5) that the emission is from an outer cone. Since N is known (and
assuming initially that we are seeing a single system, so that N=p),
we can immediately estimate the particle drift-rate from (20) and (22)
as
\eq
\lab{37}
(\Omega_D)_{out}\approx\frac{2\pi}{P}\frac{q}{4}
\en
It can be immediately seen (and is formally expressed in (24)) that
$q<{4}$ if $(\Omega_D)_{out}$ is to be less than the pulsar
rotation rate.

\subsection{The B mode}
The relatively small value of the observed $P_3$ enables us to further
constrain the possible configurations. Firstly, for either type of
cone we can see that for p=20, q=1 neither the outer cone condition
(33) nor its inner cone equivalent is met, so, in the observer's
frame, all 20 nodes cannot be linked in a single traverse round the
polar cap. Systems with an even value of q can also be excluded since
they would require an odd number of nodes. Thus q=3 remains the only
possibility if all nodes belong to the same system. However it is
theoretically possible that we are observing 2 independent but
interlocking systems of 10 nodes each, or 4 of 5 nodes etc. These
systems cannot satisfy the criterion for q=1, but are possibilities
with q=3. Nonetheless, the system's observed clockwork regularity
argues powerfully for a single integrated system, and therefore the
most likely candidate is one with p=20, q=3.

Accepting this as our working hypothesis, what characteristics will
the system possess? It will connect every third node at time interval
(from (22)) of approximately 0.212P until all 20 nodes are visited,
taking $t_c\approx{20}\times0.212=4.24$ periods for the cycle.
Assuming only that $P_3$ is in the region of 1.85, then from (30) the
net angular drift of the system after the time $t_c$ will be
approximately
\eq
\lab{38}
\Theta_c\approx\frac{76.4^o}{1.85}=41.3^o
\en
implying a net drift of about $2^o$ per node visit. The nodes are
$18^o$ apart, so node visits are separated by $3\times{18}+2=56^o$.

To make the model precise we need more exact values for both $P_3$ and
$\tau_{dz}$. Noting that the net drift angle $\Theta_c$ is close to
$40^o$, gives from (31) an estimate of $N_c$ as 8.73, close to the
integer 9. Let us therefore suppose that in reality $N_c$ is
exactly nine, giving the flow an exact repeatability. Then the further
conditions (35/36), which would apply if the flow is harmonically
coupled to the star's rotation {\em predicts} $P_3$ and $\tau_{dz}$ to
be
\begin{eqnarray}
P_3&=&\frac{1}{5}(9+\frac{1}{3})=1.86666....\\
\tau_{dz}&=&\frac{1}{5}(1+\frac{1}{3\times9})=0.2074
\end{eqnarray}
(39) is {\em precisely} the observed value of $P_3$ and is the only
value between 1.66 and 2.00 which can give harmonic coupling (these
 values corresponding to $N_c$ at 8 or 10).

From the the value of $\tau_{dz}$ in (40) we obtain an adjusted cycle
time of $t_c=20\times{0.2074}P=4.15P$, exactly one ninth of the
observed $\hat{P}_3=37.35$. With the deduced values for $P_3$ and
$\tau_{dz}$ it follows that the particle drift in the observer's
(inertial) frame has an angular speed of
\begin{eqnarray}
(\Omega_{D})_{out}&=&\frac{2\pi}{P}\times\frac{3}{20}
[\frac{1}{0.2074}+\frac{1}{3\times{1.867}}]\nonumber\\
&=&\frac{2\pi}{P}\times{0.7500}
\end{eqnarray}
so the particle drift is exactly three quarters of the star's rotation
rate. This suggests that the pulsar has adjusted both the path length
of its flow and its potential field so that this coupling occurs.

Thus we are left with a model of extraordinary harmony: $P_3$ is 9
times the basic timescale of 0.2074P, the particle drift circulates
round the magnetic axis 3 times in exactly 4 periods, requiring
$20\times{0.2074}P$ ($=4.15P$) to complete the 20-node cycle, which
then repeats itself exactly 9 times before returning to its starting
point.

In the corotating frame the picture is even simpler.  The angular
speed of the the particles $(\Omega_D)^{'}_{out}$ will now be exactly
one quarter of the pulsar rotation rate, and in the opposite
sense. The equivalent equation to (41), relating the particle drift to
the $p'$ and $q'$ seen in the corotating frame will be
\begin{eqnarray}
(\Omega_D)_{out}&=&-\frac{2\pi}{P}\times\frac{1}{20}
[\frac{1}{0.2074}+\frac{1}{1\times(3\times1.867)}]\nonumber\\
&=&-\frac{2\pi}{P}\times{0.2500}
\end{eqnarray}
It is clear that p remains unchanged from p at 20 but now $q'=1$,
whereas q=3. Thus in this frame all the nodes are visited
consecutively in a single turn, with the cycle still taking 4.15
periods, but an observer would see a much slower node drift
$P_3=3\times{1.867}=5.600$.  This finally enables us to estimate the
underlying potential difference required to support this drift.  From
(10) it follows that three quarters of the available electric
potential is screened, thus leaving a potential (from (18) with
$B_{12}=2$, P=1.1, $\frac{p'}{q'}=p'=20$) of
$\Delta{V}=5.3\times10^{12}eV$, enough to accelerate particles to
$\gamma\approx10^{7}$.

\subsection{The Q mode}
For the irregular Q-mode, which comes on abruptly and shatters the
clockwork regularity of the B-mode, an interpretation is more
difficult. But a number of features stand out. DR note that at the
outset there is a roughly 4P intensity fluctuation present, which
could be the residual of the underlying particle drift periodicity of
the 20 node system. This periodicity is never observed in the
smoothly-flowing B-mode, whose variations are those of phase rather
than intensity. We would interpret this as the particle flow between
the surface and the nodes becoming disordered, though possibly even
continuing to drift at a rate similar to that in the B-mode. The
particles now flow in `packets' rather than streams, and give rise to
the `streaky' emission which DR report.

But why does the flow become disordered? The most prominent feature in
the Q-mode power spectrum, spread across most of the pulse window,
equivalent to a periodicity of $P_3=1.293$, arises from occasional
periodic bursts against a generally chaotic background. But with p=20
this drift no longer satisfies the condition (33) for any integral
value of q. The possible suggestion is that the Q-mode has become
disordered because $P_3$ has suddenly changed to a value which
destabilises the 20-node pattern. Substituting the new $P_3$ into (30)
and (31) yield a net drift angle, $\Theta_c$, of $59.1^{o}$ and
$N_c$=6.1, yet again giving almost round figures.

Assuming therefore that the flow is attempting to achieve a regular
stable system with $N_c=6$ precisely, and taking q=2, we obtain
from (36/35) $\tau_{dz}=0.210$ and $P_3=1.300$, close to the observed but
intermittent $P_3$, but requiring a slightly larger $\tau_{dz}$. Thus
we may interpret the Q-mode as a failed attempt by the pulsar to
switch to a second harmonic state with (from (26)) a particle drift of
exactly half the pulsar rotation rate. It fails because such a drift
is not consistent with 20 nodes, and requires them to be reduced. So
why doesn't the pulsar simply reduce the number of nodes to a
compatible figure? Possibly because the current flow from and to the
polar cap, presumably proportional to the number of nodes, has to be
maintained at some fixed rate whatever the level of the electric
potential.

Note that in the catalogue of `drifting' pulsars of Rankin(1986) there
are no fewer than 6 examples of pulsars with a $P_3$ of 2.1-2.2 (the
alias of $P_3$=1.85), out of a total of 28. This seems a very high
proportion, and it is possible that some - even all - these pulsars
have exactly the same $P_3$. In addition to PSR0943+10, the catalogued
pulsars are PSR2303+30, PSR0834+06, PSR2021+50, PSR2310+42 and
PSR2020+28. PSR1933+16 has also been reported to show this phenomenon
(Oster et al., 1977, Wolszscan, 1980)) and PSR1632+24 is a clear
example (Hankins $\&$ Wolszczan, 1987). It seems there is evidence of
a pulsar subpopulation with this on-off property (see Fig 4 of Rankin
1986). Clearly aliasing bedevils a proper analysis of the drifting
behaviour of many pulsars, and more careful investigation is needed
and could provide a useful test of this model.

\section{The inclined rotator}

Many of the relatively stable (and slow-drifting) pulsars appear to
have low anges of inclination (see the examples listed in Section 5.4
and the tables of Rankin 1993b). This is unlikely to be a
coincidence. Near-axisymmetric geometry means that quantities such as
$r^{*}_{dz}$, the distance between the outer nodes and inner nodes,
and its associated time scale, $\tau^{*}_{dz}$, remain roughly
constant along the trajectory of the nodes. Thus conditions for
pair-creation will not vary too much as the nodes drift. Fig 5 shows
graphically how that, although the {\em observed} nodes circulate the
magnetic axis, the mirror nodes enclose both the magnetic and the
rotation axis. The slow drift in pulsars of this type, often observed
to be many orders of magnitude slower than the rotation period,
therefore implies that the locations of the mirror nodes are almost
precisely corotating with the dead zone surface on which they sit,
although the real particle flows which produce them will be far from
corotating.

\begin{figure}
$$\vbox{
\psfig{figure=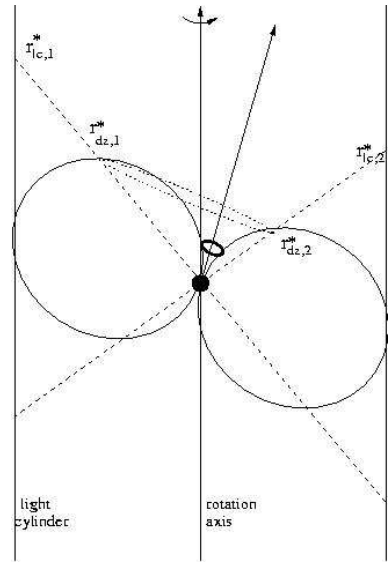,width=5.0truecm,angle=0}
}$$
\caption{The emission geometry for pulsars at a low angle of inclination.
 Note that although the rings of nodes both near the pusar surface
 (dark ring) and their mirrors on the null surface (finely dotted
 ring) encircle the magnetic axis, only the mirror ring at also
 includes the rotation axis. A low node drift (ie low subpulse
 drift-rate) therefore implies near corotation of the outer nodes on
 the surface of the corotating dead zone.}
\label{Fig5}
\end{figure}

In near-axisymmetric conditions it is easy to see how drifting
subpulses and the resulting emission cones can be maintained. But the
model here, in contrast to the RS model, requires interaction between
widely-separated regions of the magnetosphere, and it cannot be
assumed that conditions for the creation of emission cones can persist
for larger values of $\alpha$.

Figs 6, 7 and 8 show the fiducial plane which contains both the
rotation and magnetic axes for increasing angles of inclination. For
simplicity, at all angles a pure dipole geometry is assumed rather
than the more exact, but more complex, Deutsch (1956) solution for a
rotating dipole in vacuo (Arendt $\&$ Eilek, 2001).  Indeed an even
better approximation may be the complex fully-corotating field derived
analytically for inclined geometries by Beskin et al
(1993). Nonetheless our assumption has the virtue that the null
surface still intersects the plane in two straight lines (as in the
aligned case), but now makes differing angles with the dipole
axis. Here we focus attention on the changing locations of the mirror
points in this plane as $\alpha$ increases, and their possible
interactions with other features in the azimuthal plane (the plane
perpendicular to the rotation axis).

\begin{figure}
$$\vbox{
\psfig{figure=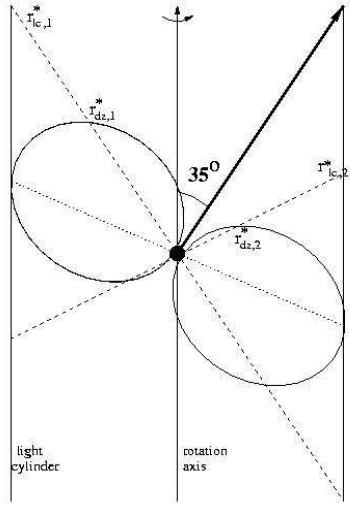,width=5.0truecm,angle=0}
}$$
\caption{Connectivity at $\alpha=35.3^o$. At this inclination the `jet' meets the
 light cylinder at the same colatitude as the light cylinder mirror
 point, enabling interaction between these regions. Furthermore the
 dead zone osculating point is capable of interaction with the
 opposite light cylinder mirror point.}
\label{Fig6}
\end{figure}

\begin{figure}
$$\vbox{
\psfig{figure=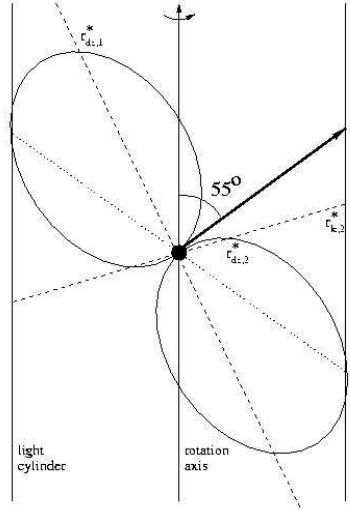,width=5.0truecm,angle=0}
}$$
\caption{Connectivity at $\alpha=54.7^o$. At this inclination the `jet' colatitude coincides
with that of the osculating point of the dead zone. For $\alpha$
larger than this the possibilities of regional interaction become
greatly reduced}
\label{fig7}
\end{figure}

\begin{figure}
$$\vbox{
\psfig{figure=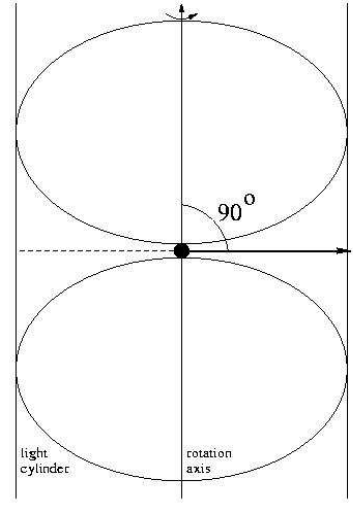,width=5.0truecm,angle=0}
}$$
\caption{Connectivity at $\alpha\simeq{90^o}$. At near-perpendicular geometries interaction
is possible between opposite poles through the opposing deadzone
osculating regions, and even through jet-to-jet contact}
\label{fig8}
\end{figure}

One such feature is a polar `jet': It is known that most pulsars
display activity in the core region of their profile, so we can
reasonably assume that a stream of particles emanates from or infalls
onto the magnetic pole along field-lines in a close bundle around the
polar field-line. If the pulsar is not aligned with its rotational
axis, this stream will cross or `impact' the light-cylinder at an
angle which for simplicity we will assume to be $\alpha$. However,
near the light-cylinder the spinning of the star will twist the polar
field-lines backwards in azimuth, together with the streaming
particles, forming a ring-like structure at this colatitude.

A second possibility is the osculating boundary of the dead zone at
the light cylinder, represented in Figs 6 and 7 by the extremes of the
dotted lines.  As is pointed out in Mestel (1999), particles
constrained to rigidly corotate close to the light-cylinder can be
expected to achieve sufficient inertia to drift across
field-lines. These particles would rotate more slowly than the rigid
rotation and form a ring around the rotation axis at the appropriate
colatitude, encountering open fieldlines and possibly contributing to
screening.

As $\alpha$ increases from zero, the angle between the rotation axis
and $r^{*}_{lc,1}$, decreases monotonically, until a point is reached
where this angle is itself equal to $\alpha$. This is at
\eq
\lab{43}
\alpha=\arctan{(\frac{1}{\sqrt{2}})}=35.3^o
\en
At this angle the `impact point' of the core jet is in the same
azimuthal plane as one of the mirror points of the inner cone (see Fig
6). In the same configuration it can be seen that other mirror point
of the inner cone (at $r^{*}_{lc,1}$) is at almost the same angle
($=2\alpha=70.6^o$) to the rotation axis as the osculating point of
the dead zone($=66.8^o$) and provides a second opportunity for
azimuthal interference, this time between the second mirror point in
this plane and, via the osculating point, the inner conal structure of
both poles.

At an inclination angle of
\eq
\lab{44}
\alpha=\arctan{\sqrt{2}}=54.7^o
\en
a further geometric coincidence occurs (Fig 7) where the magnetic axis
and the dead zone contact are at the same colatitude on opposite sides
of the rotation axis. This suggests a direct interactive link between
the polar jet and the outer cone near the light-cylinder in the
azimuthal ring perpendicular to the rotation axis. The locus of the
inner cone mirror points at the light cylinder must also cross this
ring at two places, so both cones and the core are capable of
interacting in a single system. This may account for the complex
behaviour of PSR 1237+25 (Hankins $\&$ Wright 1980), a pulsar which is
thought to have an inclination angle of $53^o$ (Rankin 1993b), and in
which the behaviour of the inner and outer cone are clearly
coordinated.

The above picture of interactivity, based only on a geometric analysis
of the relative locations of the null surface intersections and the
magnetic axis, lends a surprising degree of support to the results of
RI, in which Rankin assembles two histograms of the numbers of each
profile type versus $\alpha$ in 10-degree bins. As pointed out in
Section 2, these results require extraneous - but not unreasonable -
assumptions concerning the apparent angular size of the core region,
yet they display a dependency on $\alpha$ which strongly supports the
interactive picture developed here. The first (Fig 2 of RI) shows the
histogram for pulsars with a prominent core component (type $S_t$), a
young population with a mean age around $10^6$ yrs. This reveals a
sharp peak at $35^o$, which tails off up to $65^o$. In a further
comment (in RII) on the distribution of type $S_t$ pulsars, Rankin
finds that $S_t$ pulsars can be divided into two sub-populations, one
having no cones at any frequency and the other developing inner cones
at higher frequencies. The only factor discriminating between these
populations is that the former cluster around $35^o$ and the latter
around $50^o$.

Using our interactive model we can interpret these observations to
mean that in young pulsars with narrow light-cylinders the core
component only appears when it is possible for the core jet to
azimuthally interact, via a non-equatorial ring structure, with the
mirror regions of the inner cone, and that that it is further enhanced
by the inner cone's link to the dead zone (and hence to the outer
cone).  $\alpha=35^o$ therefore represents a `resonant' inclination
which encourages a strong jet, yet inhibits the formation of
cones. For larger $\alpha$, azimuthal interaction between the jet and
the mirror regions remains possible in planes other than the fiducial
plane, but will lack the $180^o$ symmetry as the pulsar spins. As we
approach the inclination where the jet and the dead zone can interact
directly ($\alpha=54.7^o$) an inner emission cone is able to form,
suggesting that the function of its mirror region in the particle flow
has changed now that the jet can interact with the opposite pole.

The second histogram (Fig 4 of RI) relates to the broader class of
pulsars with multiple cones or single outer cones (M and T types) and
also shows a strong peak at $35^o$, but with a somewhat wider
spread. This suggests that interaction between different regions of
the magnetosphere encourages emission cones to form for all pulsar
types. Azimuthal magnetospheric interactions with a ring structure
enabling the current to close may indeed be a prerequisite for the
`pulsar phenomenon' to appear at all. But in these generally slower
pulsars with wide light cylinders and weak jets (if any), the
phenomenon takes a different form. Although the $35^o$ peak strongly
suggests interaction with the jet region is critical, the weakness of
the jet no longer inhibits cone formation. Cone formation now depends
on a combination of the altitude of the mirror regions operating in a
non-axisymmetric geometry, the azimuthal position of the jet in
relation to the mirror regions, and almost certainly the level of
activity in the jet. The jet and the cones will exist in a symbiotic
relationship.

For $\alpha$ between $60^o$ and $85^o$ the number of pulsars of all
conal types falls off dramatically in the histograms. At these angles
the mirror points on the light-cylinder, the dead zone contact and the
jet are now widely separated from each other in colatitude,
discouraging interaction (indeed any interaction would probably be
between the opposite poles). Furthermore the field lines linking the
surface to the mirror points for the outer cones (on the surfaces of
the dead zone) are becoming severely distorted, demanding a long and
complex route across the rotation axis on the one side, and with the
distance to the mirror point ($r^{*}_{dz,2}=0.06R_{lc}$) on the other
side diminishing well below the levels demanded by the criterion of
the previous section. Similar distortions can be seen in the inner
cone mirror points, and recall the result of the previous section
that, for whatever reason, extreme high or low mirror points appear to
inhibit cone formation.

But finally, at inclinations close to $90^o$ (see Fig 8), the
situation changes dramatically and a strong peak appears near
$\alpha=90^o$ in both histograms (see again Figs 2 and 4 of RI). Now
the dead zone makes contact on both sides of the light cylinder and
fills an azimuthal ring about the rotation axis. Interaction, now
between opposite poles, is again established and pulsars of all types
are indeed found to cluster at this angle. Direct contact between the
polar jets is quite possible and and is supported by examples of
observed interpole interaction in a number of pulsars (Fowler $\&$
Wright 1982, Gil et al 1994, Biggs 1990).

Although we have compromised with reality by considering only a
strictly dipole geometry, the results of this section show that it is
possible to account for the observed dependence of pulsar type on
$\alpha$ by invoking light-cylinder interactions as the catalyst for
pulsar emission.  It would seem plausible that the observed
distribution of pulsar inclinations is not an evolutionary effect, and
that therefore many rotating magnetic neutron stars, especially at
high inclination, fail to become observable pulsars because they lack
the necessary connectivity.

\section{Theoretical Issues}
We have now arrived at the point where we must consider what physical
system might lie behind the essentially geometric picture we have built
so far.

\subsection{The global magnetosphere}

The mirror points we have identified are locations where transitional
behaviour in the charge flows of the magnetosphere might be expected:
At the light-cylinder intersection the particles will have to
negotiate the sign-change in the charge density, and also somehow keep
their velocities subluminal as they cross the light-cylinder. At the
dead zone intersection, there must be a sharp lateral transition layer
between the corotating region and a subrotating flow of charge moving
along the field-lines into and out of the gap. It is therefore not
unreasonable to suspect that these two regions may influence, or even
be responsible for, the cap emission.

The significance of these intersections was anticipated in the
axisymmetric magnetosphere model of Mestel et al. (1985, henceforth
MRWW), and similar ideas are present in that of Shibata (1995),
designed to explain gamma-ray producton in fast pulsars.  Both papers
are based on the principle that to create the torque necessary to
brake the star, the current must close outside the light-cylinder,
leaving the magnetosphere at a higher latitude and returning at a
lower.  MRWW envisage electrons streaming from the polar cap to the
light-cylinder intersection, at which point they are forced to leave
the field-lines, and postulate the creation of a gap void between this
point and the dead zone. Thus the return current must find its way
back between the outer gap and the corotating dead zone. There are
echoes of this in the configuration described earlier for
$\alpha=54.7^o$ (see Fig 7), although the axisymmetric nature of the
theoretical models inevitably precludes the possibility of using
azimuthal flow to complete the current circuit. Such connectability
within inclined rotators may turn out to be a key ingredient.

\subsection{The feedback process}

In requiring a two-way flow between the polar cap and the gap region
on both the critical fieldlines, the model here also suggests a
feedback process. The simplest model would accelerate particles from
the cap all the way to the null line without completely screening the
electric field parallel to the magnetic field lines, and would there
create pairs in an outer gap. Some positrons would then be available
to complete the feedback loop implied by the `mirror' concept and
return to the surface.  Somehow the returning positrons must interact
with the electron flow, and so on.  Models relevant to our purpose are
those which are set in conditions of near GJ screening (so-called
space charge limited flow models) and include the accelerator models
of Arons et Scharlemann (1979), Mestel $\&$ Shibata (1994), Shibata
(1997) and Mestel (1999). The attraction of these models here is that
the outflowing particles (envisioned by these authors as electrons)
are only gradually accelerated to high altitudes before Lorentz
factors of sufficient magnitude to initiate pair production are
achieved.
 
Recent promising developments of the space-charge-limited flow models
(Arons $\&$ Scharlemann 1979, HAa,b, HM, HMZ), recognising the the
role of (nonresonant) inverse Compton scattering of thermal photons -
rather than the curvature radiation operating in young pulsars (HAa,
Harding $\&$ Muslimov 2001) - as the trigger for pair production in
`slow` pulsars, demonstrate that the created pairs cannot fully screen
the electric field parallel to the magnetic field. This feature
(described in HM) fits well with the requirement here that particle
acceleration should continue up to the outergap location, and is
further supported by the critique by Jessner et al (2001) of dense
pair production models. However space charge-limited models rely
heavily on suitable axisymmetric dipole geometry to drive the
acceleration (by means of divergencies from GJ charge
densities). Observations of drifting subpulses, and the explanation
for these offered here, imply a highly non-axisymmetric potential and
charge distribution above the pole (although still with a dipolar
magnetic field). It would be of interest to examine the consequences
for these models of, say, a regular azimuthal variation in potential.

In order to create a feedback to the surface we also need a model for
pair creation in the outer magnetosphere to follow the magnetosphere
acceleration, and one which will work in `slow' pulsars, i.e. those
with periods of the order of a second.  The model of Shibata (1995) -
designed for fast pulsars - stressed that any supposed inner
accelerator system must influence the electrodynamics of the
light-cylinder region in order to successfully dispose of the star's
angular momentum. The analysis here suggests that the accelerator is
fully integrated into the dynamics of the outer magnetosphere, even
for pulsars which are not observed to emit gamma-rays, and that
pulsars have found a way of feeding back changes in the outer
magnetosphere to the polar cap. As yet, the possibility of weak,
non-cascading, intermittent, and also non-axisymmetric, pair
production in an outergap acceleration zone has not been examined, and
it therefore remains far from clear whether the outergap can sustain
itself in its highly diminished role. But the detailed work of HSa,b
(also Hirotani 2000), although focussed on high-energy pair
production, intriguingly suggests that the location of the outergap
may depend on the inflowing current from the star and hence be part of
the feedback process - thereby hinting at a mechanism for
mode-changing.

\subsection{The emission process}
In the pair-creation cascade at the mirror points the electrons will
be accelerated and injected into the wind zone beyond the
light-cylinder. But, as Shibata (1997) noted, at least a small
fraction of the positrons will trail behind the cascade process and
feel the same potential as the primary electrons and return to the
surface. As they approach the star ( e.g. Michel 1992), the positrons
will acquire a $\gamma$ factor sufficient to pair-create and an
avalanche of bunched pairs will be formed some 10 star radii above the
surface. This process is analogous to that of cosmic ray entry into
the atmosphere, as noted by Melrose (1996).  In Michel's model the
downward coherent radiation produced by this process is reflected from
the surface. The radiation is exactly radial, has the dimensions of
the polar cap (and reverses its polarisation), and would seem to be
well-suited to explain the emission features and dimensions of the
pulsar profile`s central core.  Rankin (in RI) observed that the
central (core) component of pulsar profiles scales precisely with the
opening angle of a dipole polar cap at the surface of the star, just
what would be expected from Michel's model. Effectively, the reflected
radiation will be a miniature image of the conal structure, and the
time lag behind the conal radiation at 220km will shift the core
position by around 0.7ms.  It is important to realise that the number
of returning positrons must be well below the GJ value so as not to
`poison' the outflow (Michel 1991, Lyubarskii 1992), and will impact
only a fraction of the polar cap surface. Otherwise, as Shibata et al
(1998) have pointed out, the polar cap would emit X-rays above the
observed limit formulated by Becker $\&$ Truemper(1997).

On the other hand, X-rays from the surface may be part of the
mechanism which produces pairs in the outer gap (Harding 2000, Romani
$\&$ Yadigaroglu 1995). This might explain why a cone cannot be formed
when linked to a gap above a fixed altitude, rather than to a fixed
proportion of the light-cylinder radius. Similarly, below a certain
fixed altitude pair-production, stimulated by surface X-rays, may
result in so many backflowing positrons that the flow is `poisoned'
and becomes unstable (Lyubarskii 1992).

The pairs avalanche initiated by the backflowing positrons will
clearly heat the surface immediately below it and release electrons by
a thermionic process.  These new primary electrons will have a low
$\gamma$ and must pass through the oncoming avalanche above it. If
they are to emit the coherent radiation seen in pulsar profile cones
then it must be assumed that they are bunched through instabilities in
the strictly layered avalanche. This configuration does not seem to
have been analysed in the literature, but Melrose (1996) has pointed
out that such a model avoids many of the difficulties (nothing to
maintain a pancake structure, emission angles too narrow
($\frac{1}{\gamma}$)) associated with models based on purely
outflowing streams.

\section{Conclusions}
In this paper an attempt has been made to create a feedback model for
the pulsar magnetosphere which can be directly related to
observations. It sees the magnetosphere as an integral whole, where
the polar cap and the outergap are mirrors of each other. It is
suggested that the inner and outer cones found in integrated profiles
reflect the two intersections of the Holloway null surface with the
light cylinder and the corotating dead zones respectively, and the
core components of profiles quite literally reflect the feedback from
these intersections. This simple interpretation of the profiles can
reproduce with some accuracy the observed dependence of profile
features on period and angle of inclination. The observed regularity
of profile features from pulsar to pulsar is difficult to explain by
the arbitrariness of complex surface magnetic fields.

An important feature of the model is that for the first time it
suggests an natural evolutionary link between old slower radio pulsars
and young Crab-like pulsars whose predominantly high-energy
X-ray/$\gamma$-ray emission is thought to emanate from the outergap
regions of an inclined rotating dipole (Romani $\&$ Yadigaroglu 1995).
As a pulsar ages, cascades in the outergap accelerated regions fade
(Ruderman $\&$ Cheng 1988), but a meagre yet critical pair production
still lingers, which, by interacting with polar cap production,
creates the coherent radio emission we observe in the complex radio
profiles.

The drift of particles in the entire magnetosphere often gives rise to
discrete rings of emission nodes near the polar cap. They can move in
either direction, and will have differing behaviour in different
rings. The particle flow and the observed motion of the nodes may be
chaotic, or adopt a `resonant' state where the particle circulation is
harmonically coupled to the star's rotation rate. PSR0943+10 is shown
to be in such a state, and it is argued that conditions in other stars
may permit the star to cycle through several resonant states of
stepwise increasing charge screening along open field-lines. In
general, the observed subpulse drift patterns are only the residual
drift of the particle circulation, and hence are sensitive to small
variations in the ambient electrical potential.

The angle of inclination plays an important role in determining
whether a pulsar successfully creates emission nodes and hence
observable coherent radiation. It changes the position of the null
surfaces and thereby fixes the degree to which the magnetosphere is
interconnected by means of rings of azimuthally drifting particles
around the light-cylinder. The nature of the observed dependency of
inclination on conal type suggests that the flow of particles between
the core region of the magnetic polar cap and the light-cylinder is an
integral part of the system, and this can be shown to further imply
that the two magnetic poles will generally be linked in a single
system.

A further constraint on the formation of cones, suggested by the
observed dependency of conal type on the rotational period and made
plausible by the interpretation of this model, is that emission nodes,
and their resulting integrated emission cones, can only form if the
mirror points lie between fixed altitudes of approximately 20,000km
and 70,000km.

If core emission is caused by radiation reflected from the neutron
star`s surface, then it is a prediction of the model that on close
inspection the core component will have a structure of miniature cones
(an effect possibly easier to detect in younger pulsars).

However the model leaves many theoretical questions unanswered. How
exactly can the pair creation process be made to work in the outer gap
(HSa,b), and how many positrons can really be available for the
backflow to the surface? To what degree is the proposed slow
acceleration of particles between the polar cap and the outergap
(HAa,b, HM, HMZ) affected by a returning charge flow of opposite sign
and with azimuthal dependence? Can an emission process be made to work
as slow electrons pass upwards through the descending pair avalanche?
A more radical question is whether we need particles from the surface
at all to make the radio emission. Then the model could apply to
pulsar and antipulsar alike. Finally we could speculate, as in the
global model of MRWW, that the outer ring represents a net flow of
negative charge to the surface, surrounding and balancing the positive
flow within the inner cone, and thereby solving the long-standing
current balance problem, although in sufficiently inclined pulsars
non-equatorial rings could close the current in the azimuthal plane
(see Section 7). This model has many features needing theoretical
attention, but its aim is to forge a closer link between observation
and theory, and the author will be pleased if something of this is
achieved.

\section{Acknowledgments}
The author is deeply grateful to the Astronomy Centre at the
University of Sussex for the award of a Visiting Research Fellowship,
and in particular to Professor Leon Mestel for many helpful and
enjoyable pulsar discussions. The author also thanks Dr A. Harding,
the referee, for insightful comments on the manuscript.

\section*{References}

Abrahams, A.M., Shapiro, S.L., 1991, ApJ, 374, 652\\
Arendt, P., Eilek, J.A., 2001, submitted to Ap.J.,\\
Arons, J., Scharlemann, E., 1979, ApJ, 231, 854.\\
Asseo, E., Khechinashvili, D., MNRAS, preprint\\
Becker, W., Truemper, J., 1997, $A\&A$, 326, 682.\\
Beskin, V.S., Gurevich, A.V., Istomin, Ya.N., 1993, Physics of the Pulsar Magnetosphere, 
Cambridge University Press\\
Biggs, J.D.,1990, MNRAS, 264, 341.\\
Biggs, J.D., McCulloch, P.M., Hamilton, P.A., Manchester, R.N., 1985,
 MNRAS, 215, 281.\\
Cheng, K.S., Ho, C., Ruderman, M., 1986, ApJ, 300, 500\\
Deich, W.T.S., Cordes, J.M., Hankins, T.H., Rankin, J.M., 1986, ApJ,
 300, 540.\\
Deshpande, A., Rankin, J.M., 1999, ApJ, 524, 1008.\\
Deshpande, A., Rankin, J.M., (DR) 2001, MNRAS, 322, 438\\
Deutsch, A., 1959, PASP, 68, 92.\\
Fowler, L.A., Wright, G.A.E., 1982, $A\&A$, 109, 279\\
Gil, J.A., Hankins, T.H., Nowakowski, L., 1992, in The Magnetospheric Structure and Emission
Mechanisms of Radio Pulsars, IAU Coll 128, eds T.H.Hankins, J.M.Rankin, J.A.Gil,
Pedagogical University Press, Zielona Gora, 278\\
Gil, J.A., Kijak, J., Seiradakis, J., 1993, $A\&A$, 272, 268.\\  
Gil, J.A., Jessner, A., Kijak, J., Kramer, M., Malofeev, V., Seiradakis, J.H.,
Sieber, W., Wielebinski, R., 1994, $A\&A$, 282, 45.\\ 
Gil, J.A., Krawczyk, A., 1996, MNRAS, 280, 143.\\
Gil, J.A., Sendyk, M., 2000, ApJ, 541, 351\\
Gil, J.A., Melikidze, G.I., Mitra, D., 2002a, $A\&A$, 388, 235\\
Gil, J.A., Melikidze, G.I., Mitra, D., 2002b, $A\&A$, 388, 246\\ 
Goldreich, P., Julian, W.H.,(GJ), 1969, ApJ, 157, 869.\\
Han, J.L., Manchester, R.N., 2001, MNRAS, 320, L35\\
Hankins, T.H., Wolszczan, A., 1987, ApJ, 318, 410\\
Hankins, T.H., Wright, G.A.E., 1980, Nature, 288, 681\\
Harding, A., 2000, (astr-ph/0012268)\\
Harding, A., Muslimov, A.G., 2001, ApJ, 556, 987\\
Harding, A., Muslimov, A.G., 2002, ApJ, 568, 862 (HM)\\
Harding, A., Muslimov, A.G., Zhang, B., 2002, ApJ, 576, 366 (HMZ)\\
Harding, A., Muslimov, A.G., 2003, ApJ, in press\\
Hibschmann J.A., Arons, J., 2001a, ApJ, 554, 624 (HAa)\\
Hibschmann J.A., Arons, J., 2001b, ApJ, 560, 871 (HAb)\\
Hirotani, K., 2000, MNRAS, 317, 225\\
Hirotani, K. Shibata, S., 1999, MNRAS, 308, 54 (HSa)\\
Hirotani, K. Shibata, S., 2001, MNRAS, 558, 216 (HSb)\\
Holloway, N.J., 1973, Nature Phys. Sci., 246, 6.\\ 
Jessner, A., Lesch, H., Kunzl, T., 2001, ApJ, 547, 959\\ 
Jones, P.B., 1985, Phys Rev Lett, 55, 1338\\
Jones, P.B., 1986, MNRAS, 218, 477\\
Kijak, J., 2001, MNRAS, 323, 537\\
Kijak, J., Gil, J., 1997, MNRAS, 288, 631\\
Kijak, J., Gil, J., 2003, $A\&A$, 397, 969\\
Kramer, M., Wielebinski, R., Jessner, A., Gil, J.A.,
Sieradakis, J.H., 1994, $A\&A$ Suppl., 107, 515.\\ 
Lesch, H., Jessner, A., Kramer, M., Kunzl, T., 1998, $A\&A$, 332, L21\\
Lyne, A.G., Ashworth, M., 1983, MNRAS, 204, 519.\\
Lyne, A.G., Manchester, R.N., 1988, MNRAS, 234, 477.\\
Lyubarskii, Yu.E., 1992, Astron.Astrophys, 261, 544.\\
Melrose, D.B., 1994, in Pulsars, Diamond Jubilee Symposium,
Indian Academy of Sciences, ed G.Srinivasan, 137.\\
Mestel, L., Robertson, J.A., Wang, Y.-M., Westfold, K.C., (MRWW)
1985, MNRAS, 217, 443.\\
Mestel, L., Shibata S., 1994, MNRAS, 271, 621.\\
Mestel, L., 1999, Stellar Magnetism. Oxford University Press.\\
Michel, F.C., 1991, Theory of Neutron Star Magnetospheres.
University of Chicago Press, Chicago.\\
Michel, F.C., 1992, in The Magnetospheric Structure and Emission
Mechanisms of Radio Pulsars, eds T.H.Hankins, J.M.Rankin, J.A.Gil,
Pedagogical University Press, Zielona Gora, 236\\
Mitra, D., Deshpande, A. A., 1999, Astron.Astrophys, 346, 906\\
Mitra, D., Rankin, J.M., 2002, ApJ, 577, 322\\
Neuhauser, D., Koonin, S.E., Langanke, K., 1987, Phys.Rev. A, 36, 4163\\
Nowakowski, L., 1991, ApJ, 377, 581.\\
Nowakowski, L., 1996, ApJ, 457, 868\\
Oster, L., Hilton, D.A., Sieber, W., 1977, $A\&A$, 57, 323.\\
Oster, L., Sieber, W., 1977, $A\&A$, 58, 303.\\
Petrova, S.A., 2000, $A\&A$, 360, 592.\\
Petrova, S.A., Lyubarskii, Yu.E., 2000, $A\&A$, 355, 1168.\\ 
Qiao, G.J., Xu, R.X., Liu, J.F., Han, J.L., 2000, in Pulsar
Astronomy-2000 and Beyond. IAU Colloquium 177.  Astronomical Society
of the Pacific. Eds. M.Kramer, N.Wex, R.Wielebinski, 405\\
Rankin, J.M., 1983, ApJ, 274, 333. \\
Rankin, J.M., 1986, ApJ, 301, 901.\\
Rankin, J.M., 1990 (RI), ApJ, 352, 247.\\
Rankin, J.M., 1993a (RII), ApJ, 405, 285.\\
Rankin, J.M., 1993b, ApJ.Suppl., 85, 145.\\
Romani, R.W., Yadigaroglu, I,-A., 1995, ApJ, 438, 314\\
Ruderman, M.A., Sutherland, P.G., 1975 (RS), ApJ, 196, 51.\\
Ruderman, M.A., 1976, ApJ, 203,206.\\
Seiradakis, J.H., Karastergiou A., Kramer, M., Psaltis, D., 2000, in Pulsar
Astronomy-2000 and Beyond. IAU Colloquium 177.  Astronomical Society
of the Pacific. Eds. M.Kramer, N.Wex, R.Wielebinski\\
Shibata, S., 1995, MNRAS, 276, 537.\\
Shibata, S., 1997, MNRAS, 287, 262.\\
Smith, F.G., 2003, Rep.Prog.Phys, 66, 173\\
Shibata, S.,Miyazaki, J.,Takahara, F., 1998, MNRAS, 295, L53.\\
Sturrock, P.A., 1971, ApJ, 164, 529.\\
Suleymanova, S.A., Izvekova, V.A., 1984, Sov.Astron., 28, 53.\\
Taylor, J.H., Manchester, R.N., Huguenin, G.R., 1985, ApJ, 195, 513\\
van Leeuwen, J.A., Kouwenhoven, M.L.A., Ramachandran, R., Rankin, J.M.,
Stappers, B.W., 2002, $A\&A$, 387, 169\\
van Leeuwen, J.A., Stappers, B.W., Ramachandran, R., Rankin, J.M.,
2003, $A\&A$, in press\\
Vivekanand, M., Joshi, B.C, 1997, ApJ, 477, 431\\
Wolszczan, A., 1980, $A\&A$, 86, 7.\\
Wright,G.A.E., 1978, Nature, 280, 40.\\
Wright, G.A.E., Fowler, L.A., 1981a, $A\&A$, 101, 356\\
Wright, G.A.E., Fowler, L.A., 1981b, in IAU Symp. 95, Pulsars,
 eds W.Sieber $\&$ R.Wielebinski, Dordrecht:Reidel, 221\\
\end{document}